\documentclass[11pt]{article}
\usepackage{graphicx}
\usepackage{amsmath}
\newcommand{\NN}{\hbox{I\kern-.2em\hbox{N}}}  %Naturali
\newcommand{\ZZ}{{{\rm Z}\kern-.28em{\rm Z}}} %Interi
\newcommand{\RR}{\mathop{{\rm I}\kern-.2em{\rm R}}\nolimits} %Reali
\newcommand{\QQ}{\hbox{l\kern-.36em\hbox{Q}}}  %Razionali
\newcommand{\CC}{\hbox{I\kern-.58em\hbox{C}}}

\begin{document}

\title{Detection of incompatible properties}         % Enter your title between curly braces
\author{A. Sestito\\{\scriptsize Dipartimento di Matematica, Universit\`a della Calabria, via P. Bucci 30b, 87036 Rende (Italy)} \\{\scriptsize and Istituto Nazionale Fisica Nucleare, Italy}}
      % Enter your name between curly braces
\date{}          % Enter your date or \today between curly braces
\maketitle

\abstract{In a typical two-slits experiments we face the question whether it is possible or not to attain knowledge about properties incompatible with Which-Slit property together with the  measurement of the final impact point.\\
A wide family of solutions is concretely found and an ideal experiment realizing such a detection is designed, relatively to the detection of three such incompatible properties.\\
In the case of four incompatible properties, general conditions for the existence of solutions are singled out and a particular family of solutions is provided.}

\section{Introduction}
Standard Quantum Theory \cite{vN} forbids simultaneous measurement of non-commuting observables; therefore, in a double-slit experiment it is  not generally possible to measure which-slit property and the final impact point, since these properties are represented by non-commuting operator. This notwithstanding, Which Slit knowledge can be inferred by measuring a suitable property $T$, compatible with the measurement on the final impact point. A lot of devises exploit this idea to obtain \emph{indirect} knowledge about Which Slit property (Einstein's recoiling slit \cite{Bohr}, the light-electron scattering scheme of Feymann \cite{Fey}, the micro-maser apparatus of Englert, Scully and Walther \cite{[ESW]}).\\
For the same reason, the direct measurement of the three cartesian components of a spin-$\frac{1}{2}$ particle is denied. However, in \cite{[Vaid]} (VAA) a procedure is described allowing to make inferences about  three such cartesian components. The ideal experiment proposed by VAA works as follows: one of the three non-commuting observables, $\sigma_x$, $\sigma_y$ or $\sigma_z$, is measured, on a system suitably prepared, by means of an apparatus which leaves the entire system in an eigenstate of the measured observable; after such a spin measurement, a suitable observable $A$ is measured, having the property that the outcome of the measured spin can be inferred from the outcome of $A$, without knowing which spin had been previously measured; this method yields inferences such as: if $\sigma_x$ has been measured the outcome is $+\frac{1}{2}$, if $\sigma_y$ has been measured the outcome is $-\frac{1}{2}$ and if $\sigma_z$ has been measured the outcome is $\frac{1}{2}$.\\
In \cite{[Nis]}  the problem of detecting  more than two incompatible properties is investigated on a theoretical ground, within the framework of the double-slit experiment.
Nistic\`o shows that, in a double-slit experiment, properties, incompatible with Which-slit property, can be \textit{detected} together with the knowledge of which slit each particle passes through and together with the measurement of the point of impact on the final screen. \\
This kind of detection is made possible by the fact that besides the position of the centre-mass-motion, the system possesses further degrees of freedom. As a consequence, the Hilbert space describing the entire system can be decomposed as ${\cal H}_I\otimes{\cal H}_{II}$, where ${\cal H}_I$ is the Hilbert space used to represent the observable position, and ${\cal H}_{II}$ is the Hilbert space used to represent the observable arising from the further degrees of freedom. The detection of Which Slit property $E$ is obtained by means of an observable represented by a particular projection operator $T$ acting on ${\cal H}_{II}$. The possibility of detecting an incompatible property $G$ is provided by the existence of an observable represented by another projection operator $Y$ acting on ${\cal H}_{II}$, but which can be measured together with $T$. A systematic investigation establishes \cite{[Nis]} that the existence of such an observable (projection operator) depends on the dimension of space ${\cal H}_I$.\\
A real experiment realizing such a detection is not yet performed, but finding further solutions increases the possibility of a concrete realization of such an experiment. In this perspective,
in the present work we present the results of an investigation aimed to find more general possibilities of detecting more than two incompatible properties, by using the same method proposed in \cite{[Nis]}.\\
Section 2 is devoted to the detection of three incompatible properties: Which slit property, an incompatible one and the final impact point. We translate such a problem into  mathematical terms. We treat in details the case $dim({\cal H}_1)=6$, neglecting the case $dim({\cal H}_1)<6$ which provides no solution or correlated ones \cite{[Nis]}.  A wide family of solutions is provided. We conclude the section by proposing an ideal experiment which realizes the detection at issue.\\
In section 3 the mathematical details of the derivation of the solutions proposed in section 2 is presented.\\
In section 4 the question whether two mutually incompatible properties, $G$ and $L$, both incompatible  with Which Slit property $E$, can be detected, together with  the measurement of the final impact point (four incompatible properties), is treated. In particular, we show that such a question has an affirmative answer. As in the previous case, the existence of solutions depends  on dimension of space ${\cal H}_I$; we find a particular solution for  $dim({\cal H}_1)=10$, nevertheless, in such a case the properties $L$ and $G$ turn out to be correlated.\\
The details of  the derivation of a family of this kind of solutions is presented in an appropriate section, namely section 5.

\section{Simultaneous detection of incompatible properties}
We briefly introduce the mathematical formalism to describe a two-slit experiment, allowing the detection of Which Slit property $E$ and an incompatible property, together with the final impact point.\\
The physical system consists of a localizable particle whose position observable is represented, at time $t$ in Heisenberg picture, by an operator $Q^{(t)}$ of a suitable Hilbert space ${\cal H}_I$. Let us suppose that, besides the position of the centre-of-mass motion, the system possesses further degrees of freedom, described in a different Hilbert space ${\cal H}_{II}$. As a consequence, the Hilbert space describing the entire system can be decomposed as ${\cal H}_I\otimes{\cal H}_{II}$.
Let us suppose that the Hamiltonian operator $H$ of the entire system is essentially independent of the degrees of feedom described by ${\cal H}_{II}$, so that we may assume the ideal case $H=H_I\otimes{\bf 1}_{II}$.\\
Which Slit property is a position observable, so that it is represented by a projection operator $E$ acting on  ${\cal H}_{I}$. One can make inferences about Which Slit property by means of measurements of an observable represented by  a projection operator $T$, acting on ${\cal H}_{II}$, whose outcome is correlated with the outcome of $E$. \\
Let $G$ be another property, represented by a projection operator acting on ${\cal H}_{I}$, incompatible with Which Slit property; if this new property can be detected by means of an operator $Y$ acting on ${\cal H}_{II}$ and if the detections of $T$ and $Y$ can be carried out together, then we can make simultaneous inferences about $E$ and $G$.  \\
The localization property ``the particle passes through slit 1" (which-slit property, WS) is represented in the complete Hilbert space by the operator $E=E_{I}\otimes {\bf 1}_{II}$ where $E_{I}$ is a localization operator  in ${\cal H}_I$. \\
If at time $t_1$ the particle passes through the screen supporting the  slits, we denote by $t_2$ the time of the impact on the final screen.
Given any interval $\Delta$ on the final screen  we denote by $F(\Delta)$ the projection operator representing the event ``the particle hits the final screen in a point within $\Delta$", hence concerning time $t_2$ ($F(\Delta)=F_{t_2}(\Delta)$). $F_{t_1}(\Delta)$ must have the form $F_{t_1}(\Delta)=J_{t_1}\otimes{\bf 1}_{II}$, because it is a localization operator at time $t_1$; equation $[F_{t_1}(\Delta), E]={\bf 0}$ holds but we cannot assume  $[F(\Delta), E]={\bf 0}$.
This notwithstanding, since the Hamiltonian $H$ has the form $H=H_I\otimes{\bf 1}_{II}$ then $F(\Delta)$ is a projection operator of the form $F(\Delta)=J\otimes{\bf 1}_{II}$.\\
Let  $G=G_{I}\otimes{\bf 1}_{II}$  be a projection operator representing a property incompatible with WS property $E$, i.e $[E,G]\neq 0$.\\
The possibility of detecting $G$ and $E$, together with $F(\Delta)$, is ensured by detectors $T$ and $Y$ of $E$ and $G$ respectively, of the form
$T={\bf 1}_{I}\otimes T_{II}$ and $Y={\bf 1}_{I}\otimes Y_{II}$,
satisfying $[T,Y]=0$ in such a way that $T$ and $Y$ can be measured together giving simultaneous information about  $E$ and $G$. \\
The problem at issue can be set out as follows:\\
\\
{\bf  Problem}
\emph{(P) Given the property $E=E_{I}\otimes {\bf 1}_{II}$ we want to find a  projection operator $G_{I}$  of ${\cal H}_I$, $T_{II}$ and $Y_{II}$ of ${\cal H}_{II}$, and a state vector $\Psi\in {\cal H}_I\otimes {\cal H}_{II}$ such that the following conditions are satisfied:
\begin{enumerate}
\item[(C.1)] $[E,G]\neq 0$ i.e $[E_{I},G_{I}]\neq 0$%1
\item[(C.2)] $[T,E]= 0$ and $T\Psi= E\Psi$ %4
\item[(C.3)] $[Y,G]= 0$ and $Y\Psi= G\Psi$ %5
\item[(C.4)] $[T,Y]= 0$ %7
\item[(C.5)] $\Psi\neq E\Psi\neq 0$ and $\Psi\neq G\Psi\neq 0$  %10
\item[(C.6)] $[T, F(\Delta)]=[Y, F(\Delta)]=0$
\end{enumerate}}
\noindent Condition (C.1) and equation $[T,E]= 0$ in  (C.2) express the incompatibility between properties represented by $E$ and $G$, and  compatibility between  properties represented by $T$ and $E$, respectively;
equation $T\Psi= E\Psi$ in (C.2) entails an entanglement between  $T$ and $E$, indeed it is mathematically equivalent to state that conditional probabilities satisfy
$$p(T\mid E)=\frac{\langle\Psi\mid TE\Psi\rangle}{\langle\Psi\mid E\Psi\rangle}=1=\frac{\langle\Psi\mid TE\Psi\rangle}{\langle\Psi\mid T\Psi\rangle}=p(E\mid T),
$$
in other words, outcome $1$ ($0$) for $T$ reveals the passage of the particle through slit $1$ ($2$). Condition (C.3) has a similar interpretation; (C.4)  states the simultaneous measurability of the two supplementary detections, $T$ and $Y$; condition (C.5) is added to exclude solutions corresponding to the uniteresting case  that $\Psi$ is an eigenvector  of $E$ or $G$; conditions (C.6) states that the measurement of $T$ (and $Y$) can be performed  together with the final impact point and, taking into account the form of the operators involved, it is automatically satisfied.

\subsection{Matrix  representation}
\noindent Conditions (C.1)-(C.6) can be expressed in a more useful form if the following matrix representation is adopted.\\
Let  ${(e_1,e_2,e_3,\ldots;r_1,r_2,r_3,\ldots)}$ be an orthonormal basis of  ${\cal H}_{I}$, formed by eigenvectors of $E_{I}$, such that
$E_{I}e_k=e_k$ and $E_{I}r_j=0$  for all $k$  and  $j$.
Therefore, every vector ${ \Psi}\in{\cal H}_I\otimes {\cal H}_{II}$ can be uniquely decomposed  as ${ \Psi}=\sum_j {\bf e}_{j}\otimes{\bf x}_{j}+\sum_k {\bf r}_{k}\otimes{\bf y}_{k}$, where
${\bf x}_{j}, {\bf y}_{k}\in{\cal H}_{II}$. Now,
condition (C.4) $[T,Y]=0$ implies that four projection operators, $A_i$ ($i=1,\ldots,4$), of ${\cal H}_{II}$ exist such that
$\sum_1^4 A_i={\bf 1}$,
$T_{II}=A_1+A_2$ and
$Y_{II}=A_1+A_3$.
Thereby, we choose to represent  vectors  ${\bf x}_{j}, {\bf y}_{k}\in{\cal H}_{II}$ as  column vectors
${\bf x}_{j}=
{\left({\bf a}_{j}, {\bf b}_{j}, {\bf c}_{j}, {\bf d}_{j}\right)}^t$ and
%\textrm{and }\quad
${\bf y}_{k}={\left(\boldsymbol{\alpha}_{k},\boldsymbol{\beta}_{k},\boldsymbol{\gamma}_{k},\boldsymbol{\delta}_{k}\right)}^t$, where ${\bf a}_{j}=A_1{\bf x}_{j}$, ${\bf b}_{j}=A_2{\bf x}_{j}$, ${\bf c}_{j}=A_3{\bf x}_{j}$, ${\bf d}_{j}=A_4{\bf x}_{j}$ and  $\boldsymbol{\alpha}_{j}=A_1{\bf y}_{j}$, $\boldsymbol{\beta}_{j}=A_2{\bf y}_{j}$, $\boldsymbol{\gamma}_{j}=A_3{\bf y}_{j}$, $\boldsymbol{\delta}_{j}=A_4{\bf y}_{j}$.
Then ${\Psi}\in{\cal H}$ shall be represented as a column vector
$${\Psi}=\left({\bf x}_{1}, {\bf x}_{2}, \ldots,{\bf x}_{j},\ldots; {\bf y}_{1}, {\bf y}_{2}, \ldots,{\bf y}_{k},\ldots\right)^t.$$
Once introduced suitable matrices $P=(p_{ij})$, $U=(u_{ij})$, $V=(v_{ij})$ and $Q=(q_{ij})$,we can write a linear operator $G_{I}$ of ${\cal H}_I$ in the form
Let $G_{I}={\left(
\begin{array}{cc}
 P & U \\
 V & Q
\end{array}
\right)}$; let $X$ be linear operators of  ${\cal H}_{II}$, then, according to such a representation,  any factorized linear operator $G_{I}\otimes X$ can be identified with the matrix
\begin{equation}\protect\label{eq:otimes}
G_{I}\otimes X=
{\left(
\begin{array}{cccccc}
. & . & . & . \\
p_{ij}X & . & u_{ij}X & .\\
. & . & . & . \\
v_{ij}X & . & q_{ij}X  & .\\
. & . & . & . \\
\end{array}
\right)}.
\end{equation}
Since
$E=E_{I}\otimes {\bf 1}$, $T={\bf 1}\otimes T_{II}$, $G=G_{I}\otimes {\bf 1}$ and $Y={\bf 1}\otimes Y_{II}$, then we have
$$
\begin{array}{cc}
E=
{\left(
\begin{array}{cccccccc}
 {\bf 1} & {\bf 0} & \ldots & {\bf 0} & {\bf 0} & \ldots \\
 {\bf 0} & {\bf 1} & \ldots & {\bf 0} & {\bf 0} & \ldots   \\
 \vdots & \vdots & \ddots & \vdots & \vdots & \ddots  \\
 {\bf 0} & {\bf 0} & \ldots & {\bf 0} & {\bf 0} & \ldots  \\
 {\bf 0} & {\bf 0} & \ldots & {\bf 0} & {\bf 0} & \ldots \\
 \vdots & \vdots & \ddots & \vdots & \vdots & \ddots  \\
 \end{array}
\right)}\quad
&T=
{\left(
\begin{array}{cccccc}
 {\bf T}_{II} & {\bf 0} & \ldots & {\bf 0} & {\bf 0} & \ldots \\
 {\bf 0} & {\bf T}_{II} & \ldots & {\bf 0} & {\bf 0} & \ldots   \\
 \vdots & \vdots & \ddots & \vdots & \vdots & \ddots  \\
 {\bf 0} & {\bf 0} & \ldots & {\bf T}_{II} & {\bf 0} & \ldots  \\
 {\bf 0} & {\bf 0} & \ldots & {\bf 0} & {\bf T}_{II} & \ldots \\
 \vdots & \vdots & \ddots & \vdots & \vdots & \ddots  \\
\end{array}
\right)}\\[50pt]
G={\left(
\begin{array}{cccccc}
 p_{11}{\bf 1} & p_{12}{\bf 1} & \ldots & u_{11}{\bf 1} & u_{12}{\bf 1} & \ldots \\
 p_{21}{\bf 1} & p_{22}{\bf 1} & \ldots & u_{21}{\bf 1} & u_{22}{\bf 1} & \ldots   \\
 \vdots & \vdots & \ddots & \vdots & \vdots & \ddots  \\
 v_{11}{\bf 1} & v_{12}{\bf 1} & \ldots & q_{11}{\bf 1} & q_{12}{\bf 1} & \ldots  \\
 v_{21}{\bf 1} & v_{22}{\bf 1} & \ldots & q_{21}{\bf 1} & q_{22}{\bf 1} & \ldots \\
 \vdots & \vdots & \ddots & \vdots & \vdots & \ddots  \\
\end{array}
\right)}\quad
&Y=
{\left(
\begin{array}{cccccc}
 {\bf Y}_{II} & {\bf 0} & \ldots & {\bf 0} & {\bf 0} & \ldots \\
 {\bf 0} & {\bf Y}_{II} & \ldots & {\bf 0} & {\bf 0} & \ldots   \\
 \vdots & \vdots & \ddots & \vdots & \vdots & \ddots  \\
 {\bf 0} & {\bf 0} & \ldots & {\bf Y}_{II} & {\bf 0} & \ldots  \\
 {\bf 0} & {\bf 0} & \ldots & {\bf 0} & {\bf Y}_{II} & \ldots \\
 \vdots & \vdots & \ddots & \vdots & \vdots & \ddots  \\
\end{array}
\right)}
\end{array}$$
From condition (C.2) $T\Psi=E\Psi$, we obtain  ${\bf x}_{i}=^t\left({\bf a}_i, {\bf b}_i, {\bf 0}, {\bf 0}\right)$ and ${\bf y_j}=^t\left({\bf 0},{\bf 0},  \boldsymbol{\gamma}_j, \boldsymbol{\delta}_j\right)$.
Hence, condition (C.3) $Y\Psi=G\Psi$ is equivalent to
%S
\begin{equation}\protect\label{eq:S}
\begin{array}{cccccccc}
&(i-A)& \left\{ \begin{array}{ll}
\sum_i p_{ji}{\bf a}_i={\bf a}_j \\
\sum_i p_{ji}{\bf b}_i={\bf 0}
\end{array} \right.
\qquad
&(ii-B)&\left\{ \begin{array}{ll}
\sum_l u_{jl} \boldsymbol{\gamma}_l={\bf 0}\\
\sum_l u_{jl}\boldsymbol{\delta}_l={\bf 0}
\end{array} \right.\\
\\[10pt]
&(iii-C)&
\left\{ \begin{array}{ll}
\sum_i v_{ki}{\bf a}_i={\bf 0} \\
\sum_i v_{ki}{\bf b}_i={\bf 0}
\end{array} \right.
\qquad
&(iv-D)&
\left\{ \begin{array}{ll}
\sum_l q_{kl} \boldsymbol{\gamma}_l=\boldsymbol{\gamma}_k\\
\sum_l q_{kl}\boldsymbol{\delta}_l={\bf 0}
\end{array} \right.
\end{array}\end{equation}

\subsection{Solutions of problem \emph{(P)}}
So far we have established general constraints to be satisfied by every solution of the problem, independently of  ranks of $E$, $G$ and $A_i$, with $i=1,\ldots,4$, and therefore of  dimensions of space ${\cal H}_{I}$ and ${\cal H}_{II}$.\\ We restrict the search to the case that the two slit are symmetrical: this leads to exclude odd dimension of ${\cal H}_{I}$ and, moreover, to assume that $rank(L)=rank(I-L)=dim({\cal H}_{I})/2$.\\
In the rest of this section, we present the results of an investigation aimed to find  general families of solutions, whose existence, according to \cite{[Nis]}, depends on  dimension of ${\cal H}_{I}$.
The detailed analysis of the derivation of such solutions is a rather technical matter; so, they are displaced in the next section;  here we  only sketch the idea.\\
From (\protect\ref{eq:S}), by using basic notions of linear algebra, it can be shown that, in order  to attain non-correlated solutions, the triples of component-vectors, $\lbrace \boldsymbol{\delta}_1, \boldsymbol{\delta}_2, \boldsymbol{\delta}_3\rbrace$, $\lbrace \boldsymbol{\gamma}_1, \boldsymbol{\gamma}_2, \boldsymbol{\gamma}_3\rbrace$, $\lbrace {\bf b}_1, {\bf b}_2, {\bf b}_3\rbrace$, and $\lbrace \boldsymbol{\delta}_1, \boldsymbol{\delta}_2, \boldsymbol{\delta}_3\rbrace$, are generated by just one vector. Briefly,  (ii-B) and $U\neq {\bf 0}$ imply one of the three vectors ${\bf y}_1$, ${\bf y}_2$, ${\bf y}_3$, say ${\bf y}_1$, is a linear combination of the remaining two
\begin{equation}\protect\label{eq:dependence}
\left\{\begin{array}{c}
\boldsymbol{\gamma}_1=\lambda_2\boldsymbol{\gamma}_2+\lambda_3\boldsymbol{\gamma}_3\\
\boldsymbol{\delta}_1=\lambda_2\boldsymbol{\delta}_2+\lambda_3\boldsymbol{\delta}_3
\end{array}
\right.
\end{equation}
In system (iv-D) we get
\begin{equation}\protect\label{eq:q1}
\left\{
\begin{array}{ll}
(q_{j1}\lambda_2+q_{j2})\boldsymbol{\gamma}_2+(q_{j1}\lambda_3+q_{j3}){\bf\boldsymbol{\gamma}}_3={\bf\boldsymbol{\gamma}}_j\\[10pt]
(q_{j1}\lambda_2+q_{j2})\boldsymbol{\delta}_2+(q_{j1}\lambda_3+q_{j3})\boldsymbol{\delta}_3={\bf 0}
\end{array}
\right.
\end{equation}
If vectors $\boldsymbol{\delta}_2$ and $\boldsymbol{\delta}_3$ are linearly independent, then second equation in (\ref{eq:q1}) implies $(q_{j1}\lambda_2+q_{j2})=(q_{j1}\lambda_3+q_{j3})=0$, so that first equation in (\ref{eq:q1}) yields $\boldsymbol{\gamma}_j={\bf 0}$ for all $j$, hence ${\bf y}_j={({\bf 0},{\bf 0},{\bf 0},\boldsymbol{\delta}_j)}^t$;
in a similar way, ${\bf b}_1$ and ${\bf b}_2$ linearly independent imply ${\bf x}_j={({\bf 0},{\bf b}_j,{\bf 0},{\bf 0})}^t$. If a solution of \emph{(P)} exists, then $G\Psi=Y\Psi=0$ (meaningless solution). If a solution exists such that $\boldsymbol{\delta}_2,\boldsymbol{\delta}_3$ are linearly independent and
  ${\bf b}_2$, ${\bf b}_3$ linearly dependent, then ${\bf x}_j={({\bf a}_j,{\bf b}_j,{\bf 0},{\bf 0})}^t$
and ${\bf y}_j={({\bf 0},{\bf 0},{\bf 0},\boldsymbol{\delta}_j)}^t$; in this case equation $YT\Psi=Y\Psi$ holds, which is equivalent to say that conditional probabilities satisfy
$$p(T|Y)=\frac{\langle\Psi|T\Psi\rangle}{\langle\Psi\mid Y\Psi\rangle}=1$$
and this means that each time a particle is sorted by $T$, then it is certainly sorted by $Y$; therefore for all aventual solution, property $G$ must be correlated with Which-Slit property $E$.\\
Using equations (\protect\ref{eq:dependence}) in  (iv-D) we have that non-correlated solutions exist if and only if $\boldsymbol{\delta}_2$ and $\boldsymbol{\delta}_3$ (${\bf b}_2$ and ${\bf b}_3$) are linearly dependent, say $\boldsymbol{\delta}_3=\lambda\boldsymbol{\delta}_2$ (${\bf b}_3=\mu {\bf b}_2$); thereby, we find $(u_{j1}\lambda_2+u_{j2})=-\lambda(u_{j1}\lambda_3+u_{j3})=0$. Moreover, (iv-D) implies $\boldsymbol{\gamma}_2$ and $\boldsymbol{\gamma}_3$ can be neither linearly independent nor dependent as $\boldsymbol{\gamma}_3=\lambda\boldsymbol{\gamma}_2$; a  constant $\lambda_4\neq\lambda$ must exists such that $\boldsymbol{\gamma}_3=\lambda_4\boldsymbol{\gamma}_2$ (and similarly ${\bf a}_3=l_4{\bf a}_2$).\\
As a consequence we also attain  the form $Q$ and $U$ have to do in order to satisfy (ii-B) and (iv-D); nevertheless, self-adjointness of $Q$ yields to rather difficult calculation; hence we prefer make easier the search with the choice $\boldsymbol{\gamma}_2={\bf 0}$.\\
Among general solutions of (\ref{eq:S}), we select those which satisfy ${G_I}^{\star}={G_I}$ only;
we get
\begin{equation*}
%\begin{array}{c}
\protect\label{eq:G} P={\left(
\begin{array}{ccc}
p &-\mu_2 \left(p-\frac{{|\mu_3|}^2}{1+{|\mu_3|}^2}\right) &\mu_3(1-p)\\[10pt]
-\overline{\mu}_2\left(p-\frac{|{\mu_3|}^2}{1+{|\mu_3|}^2}\right)
& |\mu_2|^2
\left(p-\frac{{|\mu_3|}^2}{1+{|\mu_3|}^2}\right) &\mu_3\overline{\mu}_2\left(p-\frac{{|\mu_3|}^2}{1+{|\mu_3|}^2}\right)\\[10pt]
\overline{\mu}_3(1-p) &
\overline{\mu}_3\mu_2\left(p-\frac{{|\mu_3|}^2}{1+{|\mu_3|}^2}\right)
&1-{|\mu_3|}^2(1-p)
\end{array}
\right),}\end{equation*}%\\[50pt]
$$Q={\left(
\begin{array}{ccc}
q &-\lambda_2\left(q-\frac{{|\lambda_3|}^2}{1+{|\lambda_3|}^2}\right) &\lambda_3(1-q)\\[10pt]
-\overline{\lambda}_2\left(q-\frac{{|\lambda_3|}^2}{1+{|\lambda_3|}^2}\right)
&|\lambda_2|^2\left(q-\frac{{|\lambda_3|}^2}{1+{|\lambda_3|}^2}\right)
&\lambda_3\overline{\lambda}_2\left(q-\frac{{|\lambda_3|}^2}{1+{|\lambda_3|}^2}\right)\\[10pt]
\overline{\lambda}_3(1-q)
&\overline{\lambda}_3\lambda_2\left(q-\frac{{|\lambda_3|}^2}{1+{|\lambda_3|}^2}\right)
&1-{|\lambda_3|}^2(1-q)
\end{array}
\right),}$$\\[5pt]
$$U={\left(
\begin{array}{ccc}
u &-{\lambda_2}u &-{\lambda_3}u\\[10pt]
-\overline{\mu}_2u &\lambda_2\overline{\mu}_2u &\lambda_3\overline{\mu}_2u\\[10pt]
-\overline{\mu}_3u &\lambda_2\overline{\mu}_3u
&\lambda_3\overline{\mu}_3u
\end{array}
\right),}$$ $$\Psi={({\bf x}_1,{\bf x}_2,{\bf x}_3;{\bf y}_1,{\bf
y}_2,{\bf y}_3)}^t,$$ where
$$\begin{array}{ll}
{\bf x}_1={(\mu_3{\bf a}_3,\frac{\mu_2}{{|\mu_3|}^2+1}{\bf
b}_2,{\bf 0},{\bf 0})}^t &\qquad
{\bf y}_1={({\bf 0},{\bf 0},\lambda_3\boldsymbol{\gamma}_3,\frac{\lambda_2}{{|\lambda_3|}^2+1}\boldsymbol{\delta_2},{\bf 0})}^t\\
{\bf x}_2={({\bf 0},{\bf b}_2,{\bf 0},{\bf 0})}^t &\qquad
{\bf y}_2={({\bf 0},{\bf 0},{\bf 0},\boldsymbol{\delta_2})}^t\\
{\bf x}_3={({\bf
a}_3,-\frac{\mu_2\overline{\mu}_3}{{|\mu_3|}^2+1}{\bf b}_2,{\bf
0},{\bf 0})}^t &\qquad {\bf y}_3={({\bf 0},{\bf
0},\boldsymbol{\gamma}_3,-\frac{\lambda_2\overline{\lambda}_3}{{|\lambda_3|}^2+1}\boldsymbol{\delta_2})}^t
\end{array}$$
\noindent Self-adjointness of $G_I$ implies that
$V={\overline{U}}^t$, moreover
$\lambda=-\frac{\lambda_2\overline{\lambda}_3}{1+|\lambda_3|^2}$
and $\mu=-\frac{\mu_2\overline{\mu}_3}{1+|\mu_3|^2}$.\par Such a
solution completely solves the problem if numbers $p,u,q\neq 0$
can be chosen in such a manner that $G$ turns out to be
idempotent. It is easily shown that idempotence implies
\begin{itemize}
\item[1.] $\frac{{|\mu_3|}^2}{1+{|\mu_3|}^2}<p<\frac{{|\mu_3|}^2}{1+{|\mu_3|}^2}+\frac{1}{1+|\mu_2|^2+{|\mu_3|}^2}$,
\item[2.] $u=e^{i\theta}\sqrt{\frac{\left( p-\frac{{|\mu_3|}^2}{1+{|\mu_3|}^2}\right)-
\left(1+|\mu_2|^2+{|\mu_3|}^2\right)\left(p-\frac{{|\mu_3|}^2}{1+{|\mu_3|}^2}\right)^2}{1+|\lambda_2|^2+|\lambda_3|^2}}$,
where $\theta$ is a real number,
\item[3.] $q=\frac{1-\left(1+|\mu_2|^2+{|\mu_3|}^2\right)\left(p-\frac{{|\mu_3|}^2}{1+{|\mu_3|}^2}\right)}{1+|\lambda_2|^2+|\lambda_3|^2}$.
\end{itemize}
Therefore for every  real number
$\frac{{|\mu_3|}^2}{1+{|\mu_3|}^2}<p<\frac{{|\mu_3|}^2}{1+{|\mu_3|}^2}+\frac{1}{1+|\mu_2|^2+{|\mu_3|}^2}$,
every $\theta\in{\RR}$ and every
$\mu_2,\mu_3,\lambda_2,\lambda_3\in{\CC}$ we have a solution  of
(${\cal P}$). \\
No constraint is imposed to $rank(G_I)$, i.e.  to the trace  of the projection operator $G$, hence these parameters are not all independent. \\
For every choice we attain a solution, ${G_I}^1, {G_I}^2, \ldots$; then we can state there are several properties  $G^1, G^2,\ldots$ incompatible with Which-Slit property $E$ but detectable together with it. However, we notice that, taking into account (C.3), every $G^i$ transforms $\Psi$ in $Y\Psi$.\\
Our solutions form a rather wide family; however, it
is not exhaustive, because of the choice $\boldsymbol{\gamma}_2={\bf 0}$; if the case $\boldsymbol{\gamma}_2\neq {\bf 0}$ is taken into account, the problem would be completely solved. The family singled out by Nistic\`o in \cite{[Nis]} is just a subfamily of the present one corresponding to the particular choice $\mu_3=\lambda_3=0$ and $\lambda_2=\mu_2=1$.  \\
In next subsection an ideal experiment, not concretely performable, that realizes the detection at issue, is proposed.

\subsection{\bf Ideal experiment}
\noindent
Until now the treatment has been carried out on a theoretical ground only. Now we describe an ideal apparatus following the results of the previous sections.\\
The experimental set-up corresponds to the particular solution with parameters $\mu_2=\lambda_2=\sqrt 3$ and $\mu_3=\lambda_3=1$.\\
The system consists of an electrically neutral particle of  spin $\frac{3}{2} $; the position of its centre-of-mass is described in  space ${\cal H}_I$. The further degrees of freedom, described in ${\cal H}_{II}$, concerns the spin of the particle. \\
Let us suppose that, after crossing the screen with the slits, each particle passes through a non-uniform  magnetic field, with gradient along the direction $z$ (fig.1).
\begin{figure}[h]
\protect\label{figura}
\begin{center}
\includegraphics[height=4cm]{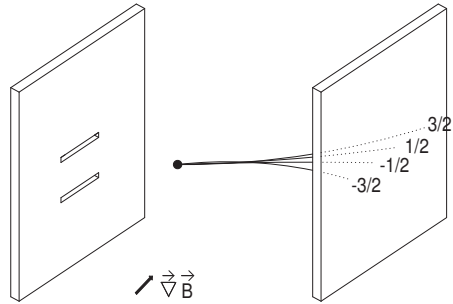}
\caption{Experimental set-up}
\end{center}
\end{figure}
The beam splits into four beams and the deflection of each particle depend of the component of the spin in the direction of the magnetic field gradient. Hence the measurement of the amount of deflection of the particle  indicates the value of its spin component.\\
Let $A$  be the projection operator representing the event ``the spin component in the z-direction is $\frac{3}{2}$". Similarly we define operators $B$, $C$ and $D$ associated to the spin components $\frac{1}{2}$, $-\frac{1}{2}$ and $-\frac{3}{2}$, respectively. We denote their respective eigenvectors relative to the eigenvalue 1 by $\mid\frac{3}{2}\rangle$, $\mid\frac{1}{2}\rangle$, $\mid-\frac{1}{2}\rangle$ and $\mid-\frac{3}{2}\rangle$.
By $\psi_i$, $i=1,\ldots,6$ we denote orthonormal eigenfunctions of ${\cal H}_1$ such that $\psi_1$, $\psi_2$, $\psi_3$ lie in $E_I{\cal H}_1$ and $\psi_4$, $\psi_5$, $\psi_6$ lie in $\left(I-E_I\right){\cal H}_1$.
Let the state vector of the entire system be
{\setlength\arraycolsep{2pt}
\begin{eqnarray*}
{\psi}&=&\frac{1}{3}\left\{\left(\psi_1+\psi_2\right)\mid\frac{3}{2}\rangle+\left(\frac{\sqrt 3}{2}\psi_1+\psi_2-\frac{\sqrt 3}{2}\psi_3\right)\mid\frac{1}{2}\rangle\right\}+{}\\
&{}+&\frac{1}{3}\left\{\left(\psi_4+\psi_6\right)\mid-\frac{1}{2}\rangle+\left(\frac{\sqrt 3}{2}\psi_4+\psi_5-\frac{\sqrt 3}{2}\psi_6\right)\mid-\frac{3}{2}\rangle\right\},
\end{eqnarray*}}
which, within our representation, coincides with
$${\Psi}=\frac{1}{3}{\left(1, \frac{\sqrt 3}{2},0,0,1,1,0,0,0,-\frac{\sqrt 3}{2},0,0;0,0,1,\frac{\sqrt 3}{2},0,0,0,1,0,0,1,-\frac{\sqrt 3}{2}\right)}^t.$$
According with the results of previous section,
with respect to this state vector there exists
a Which-Slit detector $T=I\otimes(A+B)$ and a detector $Y=I\otimes(A+C)$ of a property $G=G_I\otimes I$, incompatible with property $E$; $G_I$, with  the particular choice, is the projection operator
$$G_I=\left(
\begin{array}{cccccc}
 \frac{2}{3} & -\frac{1}{2 \sqrt{3}} & \frac{1}{3} & \frac{1}{6 \sqrt{5}} & -\frac{1}{2 \sqrt{15}} & -\frac{1}{6 \sqrt{5}} \\[10pt]
 -\frac{1}{2 \sqrt{3}} & \frac{1}{2} & \frac{1}{2 \sqrt{3}} & -\frac{1}{2 \sqrt{15}} & \frac{1}{2 \sqrt{5}} & \frac{1}{2 \sqrt{15}} \\[10pt]
 \frac{1}{3} & \frac{1}{2 \sqrt{3}} & \frac{2}{3} & -\frac{1}{6 \sqrt{5}} & \frac{1}{2 \sqrt{15}} & \frac{1}{6 \sqrt{5}} \\[10pt]
 \frac{1}{6 \sqrt{5}} & -\frac{1}{2 \sqrt{15}} & -\frac{1}{6 \sqrt{5}} & \frac{8}{15} & -\frac{1}{10 \sqrt{3}} & \frac{7}{15} \\[10pt]
 -\frac{1}{2 \sqrt{15}} & \frac{1}{2 \sqrt{5}} & \frac{1}{2 \sqrt{15}} & -\frac{1}{10 \sqrt{3}} & \frac{1}{10} & \frac{1}{10 \sqrt{3}} \\[10pt]
 -\frac{1}{6 \sqrt{5}} & \frac{1}{2 \sqrt{15}} & \frac{1}{6 \sqrt{5}} & \frac{7}{15} & \frac{1}{10 \sqrt{3}} & \frac{8}{15}
\end{array}
\right)
$$
Therefore, the ideal experiment just described allows to make inferences about three non-commuting observables: the position of the final impact point is inferred from a direct measurement of $F(\Delta)$; Which-Slit property $E$ is inferred by the outcome of detector $T$ and  property $G$ is inferred by the outcome of detector $Y$.\\
We have to notice the difference between the meaning of our results with respect to that obtained by VAA. According to this latter, from the outcome of $A$, the outcome of the performed spin-component measurement can be retrodicted; while the remaining inferences cannot be considered as detections; furthermore inferences can be drawn only under the hypothesis that the spin measurement actually performed leaves the system in an eigenstate of the measured observable.\\
We stress the ideal character of the experiment just described.
In order to make it meaningful, we would be able to identify the observable represented by $G$ by a physical point of view. Nevertheless,
practical difficulties of creating the initial entangled state $\Psi$ are the real obstacles in realizing the designed experiment . Hence,  even if a real experiment for simultaneous detection of Which Slit property, an incompatible one and the final impact point is not yet performed, a wider family of solutions is a contribution to increase the possibility of a concrete realization.

\section{Derivation of a family of solutions}\protect\label{subsec:(3p)Derivation of a family of
solutions} \noindent This section is devoted to find a  detailed
derivation of the family of solutions presented in previous
section.\par We are seeking for solutions such that the rank of
$L$ is 3, so that $i$, $j$, $k$, $l$ take values in $\lbrace
1,2,3\rbrace$. No constraint is imposed to the ranks of $A_i$,
with $i=1,2,3,4$, and hence to the dimension of ${\cal H}_{II}$.
If ${\cal H}_{I}=6$, then $\Psi={({\bf x}_1,{\bf x}_2,{\bf
x}_3;{\bf y}_1,{\bf y}_2,{\bf y}_3)}^t$, so that $P$, $U$, $V$ and
$Q$ are $3\times 3$ matrices. Since $U\neq {\bf 0}$, (ii-B)
implies that  one of the three vectors , ${\bf y}_2$, ${\bf y}_3$
is linear combination of the remaining two, say ${\bf y}_1$, so
that complex numbers $\lambda_2$, $\lambda_3$ must exist such that
\begin{equation}
\left\{\begin{array}{ll}
\boldsymbol{\gamma}_1=\lambda_2\boldsymbol{\gamma}_2+\lambda_3\boldsymbol{\gamma}_3\\[10pt]
\boldsymbol{\delta}_1=\lambda_2\boldsymbol{\delta}_2+\lambda_3\boldsymbol{\gamma}_3
\end{array}
\right.
\end{equation}
In system (iv-D) we get
\begin{equation}\protect\label{eq:q1}
\left\{
\begin{array}{ll}
(q_{j1}\lambda_2+q_{j2})\boldsymbol{\gamma}_2+(q_{j1}\lambda_3+q_{j3}){\bf\boldsymbol{\gamma}}_3={\bf\boldsymbol{\gamma}}_j\\[10pt]
(q_{j1}\lambda_2+q_{j2})\boldsymbol{\delta}_2+(q_{j1}\lambda_3+q_{j3})\boldsymbol{\delta}_3={\bf
0}
\end{array}
\right.
\end{equation}
If vectors $\boldsymbol{\delta}_2$ and $\boldsymbol{\delta}_3$ are
linearly independent, then second equation in (\ref{eq:q1})
implies $(q_{j1}\lambda_2+q_{j2})=(q_{j1}\lambda_3+q_{j3})=0$, so
that first equation in (\ref{eq:q1}) yields
$\boldsymbol{\gamma}_j={\bf 0}$ for all $j$, hence ${\bf
y}_j=^t({\bf 0},{\bf 0},{\bf 0},\boldsymbol{\delta}_j)$.
In a similar way, ${\bf b}_1$ and ${\bf b}_2$ linearly independent imply ${\bf x}_j=^t({\bf 0},{\bf b}_j,{\bf 0},{\bf 0})$. If a solution of \emph{(P)} exists, then $G\Psi=Y\Psi=0$.  A detailed analysis in \cite{[Nis]} shows that the only case which can lead  meaningful solutions without correlation is  ${\bf b}_2$, ${\bf b}_3$ linearly dependent and $\boldsymbol{\delta}_2$, $\boldsymbol{\delta}_3$ linearly dependent. \\
Let us suppose
$\boldsymbol{\delta}_3=\lambda\boldsymbol{\delta}_2$ with
$\lambda\neq 0$ (and similarly ${\bf b}_3=\mu {\bf b}_2$ with
$\mu\neq 0$). Thereby, in (ii-B) we get
\begin{equation}\protect\label{eq:u1}
    \left\{\begin{array}{ll}
          (u_{j1}\lambda_2+u_{j2})\boldsymbol{\gamma}_2+(u_{j1}\lambda_3+u_{j3})\boldsymbol{\gamma}_3={\bf
                            0}\cr
          \lbrack(u_{j1}\lambda_2+u_{j2})+\lambda(u_{j1}\lambda_3+u_{j3})\rbrack\boldsymbol{\delta}_2={\bf 0}.
    \end{array}\right.
\end{equation}
If $\boldsymbol{\delta}_2={\bf 0}$ then  second equation in
(\ref{eq:u1}) is satisfied and $\boldsymbol{\delta}_j={\bf 0}$ for
all $j$, so that ${\bf y}_j=^t({\bf 0},{\bf
0},\boldsymbol{\gamma}_j,{\bf 0})$. In a similar way, ${\bf
b}_2={\bf 0}$ implies ${\bf b}_j={\bf 0}$ for all $j$, so that
${\bf x}_j=^t({\bf a}_j,{\bf 0},{\bf 0},{\bf 0})$. Given a state
$\Psi$, we obtain the following implications:
\begin{itemize}
\item[1.] ${\bf b}_2={\bf 0}$ and $\boldsymbol{\delta}_2={\bf 0}$  imply
          ${\bf x}_j=^t({\bf a}_j,{\bf 0},{\bf 0},{\bf 0})$ and ${\bf
          y}_j=^t({\bf 0},{\bf 0},\boldsymbol{\gamma}_j,{\bf 0})$; in this
          case if a solution exists, then $G\Psi=\Psi$. Therefore meaningful
          solutions cannot exist;
\item[2.] ${\bf b}_2\neq{\bf 0}$ and $\boldsymbol{\delta}_2={\bf 0}$  imply
          ${\bf x}_j=^t({\bf a}_j,{\bf b}_j,{\bf 0},{\bf 0})$ and ${\bf y}_j=^t({\bf 0},{\bf 0},\boldsymbol{\gamma}_j,{\bf 0})$;
          if a solution exists, then $T'\Psi=T'Y'\Psi$, that is to say property $G$ must be correlated with WS property $E$;
\item[3.] ${\bf b}_2={\bf 0}$ and $\boldsymbol{\delta}_2\neq{\bf 0}$  imply
          ${\bf x}_j=^t({\bf a}_j,{\bf 0},{\bf 0},{\bf 0})$ and ${\bf
          y}_j=^t({\bf 0},{\bf
          0},\boldsymbol{\gamma}_j,\boldsymbol{\delta}_j)$; if a solution
          exists, then $TY\Psi=T\Psi$. As in the previous case property $G$ must
          be correlated with WS property $E$;
\item[4.] ${\bf b}_2\neq{\bf 0}$ and $\boldsymbol{\delta}_2\neq{\bf 0}$  imply
          ${\bf x}_j=^t({\bf a}_j,{\bf b}_j,{\bf 0},{\bf 0})$ and ${\bf y}_j=^t({\bf 0},{\bf 0},\boldsymbol{\gamma}_j,\boldsymbol{\delta}_j)$;
          this is the only case that can lead to solution without correlation.
\end{itemize}
Hence, we are interested  only in case (4).\par Since
$\boldsymbol{\delta}_2\neq{\bf 0}$, second equation in
(\ref{eq:u1}) is satisfied if and only if
\begin{equation}\protect\label{eq:u2}
    (u_{j1}\lambda_2+u_{j2})=-\lambda(u_{j1}\lambda_3+u_{j3})
\end{equation}
so that first equation in (\ref{eq:u1}) becomes
\begin{equation}\protect\label{eq:u3}
    (u_{j1}\lambda_3+u_{j3})(\boldsymbol{\gamma}_3-\lambda\boldsymbol{\gamma}_2)={\bf 0}.
\end{equation}
If we suppose
$\boldsymbol{\gamma}_3=\lambda\boldsymbol{\gamma}_2$, then in
(\ref{eq:q1}) we get
\begin{equation}\protect\label{eq:qq}
    \left\{\begin{array}{ll}
           \lbrack(q_{j1}\lambda_2+q_{j2})+\lambda(q_{j1}\lambda_3+q_{j3})\rbrack\boldsymbol{\gamma}_2=\boldsymbol{\gamma}_j\cr
           \lbrack(q_{j1}\lambda_2+q_{j2})+\lambda(q_{j1}\lambda_3+q_{j3})\rbrack\boldsymbol{\delta}_2={\bf
           0}.
    \end{array}\right.
\end{equation}
Since $\boldsymbol{\delta}_2\neq{\bf 0}$, second equation of
(\protect\ref{eq:qq}) implies that
 $(q_{j1}\lambda_2+q_{j2})=-\lambda(q_{j1}\lambda_3+q_{j3})$; hence $\boldsymbol{\gamma}_j={\bf 0}$, for all $j$, follows from the first equation in
 (\protect\ref{eq:qq}). Analogous reasoning for equations in (i-A)
 leads to ${\bf a}_j={\bf 0}$, for all $j$; hence%from conditions
% (\protect\ref{eq:TPsi}) and (\protect\ref{eq:YPsi})
, if a
 solution exists, it corresponds to the uninteresting case $G\Psi={\bf
 0}$.
 %Therefore for all eventual solutions corresponding to both cases, property $G$ must be correlated with WS
 %property $E$. So  we conclude that (\protect\ref{eq:u3})  yields
%\begin{equation}\protect\label{eq:u4}
 %     (u_{j1}\lambda_3+u_{j3})=0.
%\end{equation}
\par
If in (\protect\ref{eq:u3})  we consider all possibilities for
$\boldsymbol{\gamma}_1$ and $\boldsymbol{\gamma}_2$, we get:
\begin{itemize}
    \item[a.] $\boldsymbol{\gamma}_3=\lambda\boldsymbol{\gamma}_2$,
    \item[b.] $\boldsymbol{\gamma}_2$ and $\boldsymbol{\gamma}_3$ linearly independent,
    \item[c.] $\boldsymbol{\gamma}_3=\lambda_4\boldsymbol{\gamma}_2$,
    \item[d.] $\boldsymbol{\gamma}_2={\bf 0}$ and $\boldsymbol{\gamma}_3\neq{\bf 0}$.
\end{itemize}
Now we draw the consequences of (b), (c) and (d), since in case
(a) eventual solutions are uninteresting.
\par
\vspace{15pt} \centerline{\bf Case (b).} \vspace{10pt}\noindent If
vectors $\boldsymbol{\gamma}_2$ and $\boldsymbol{\gamma}_3$ are
linear independent, then (\ref{eq:q1}) becomes
\begin{equation}\protect\label{eq:q3}
      \left\{\begin{array}{c}
            (q_{j1}\lambda_2+q_{j2})\boldsymbol{\gamma}_2+(q_{j1}\lambda_3+q_{j3})\boldsymbol{\gamma}_3=\boldsymbol{\gamma}_j\cr
            \lbrack(q_{j1}\lambda_2+q_{j2})+\lambda(q_{j1}\lambda_3+q_{j3})\rbrack\boldsymbol{\delta}_2={\bf 0}.
      \end{array}\right.
\end{equation}
Since $\boldsymbol{\delta}_2\neq {\bf 0}$, then second equation in
(\ref{eq:q3}) implies
$(q_{j1}\lambda_2+q_{j2})=-\lambda(q_{j1}\lambda_3+q_{j3})$, so
that first equation in  (\ref{eq:q3}) yields
$(q_{j1}\lambda_3+q_{j3})(\boldsymbol{\gamma}_3-\lambda\boldsymbol{\gamma}_2)=\boldsymbol{\gamma}_j$.
Using this relation we get
\begin{equation}\protect\label{eq:q4}
       \left\{\begin{array}{lll}
           (q_{11}\lambda_3+q_{13})(\boldsymbol{\gamma}_3-\lambda\boldsymbol{\gamma}_2)=\lambda_2\boldsymbol{\gamma}_2+
                                \lambda_3\boldsymbol{\gamma}_3\cr
           (q_{21}\lambda_3+q_{23})(\boldsymbol{\gamma}_3-\lambda\boldsymbol{\gamma}_2)=\boldsymbol{\gamma}_2\cr
           (q_{31}\lambda_3+q_{33})(\boldsymbol{\gamma}_3-\lambda\boldsymbol{\gamma}_2)=\boldsymbol{\gamma}_3.
       \end{array}\right.
\end{equation}
Second equation in (\ref{eq:q4}) implies
\begin{equation}\protect\label{eq:q5}
       \left\{\begin{array}{ll}
              (q_{21}\lambda_3+q_{23})=0\cr
               -\lambda(q_{21}\lambda_3+q_{23})=1.
       \end{array}\right.
\end{equation}
Then system (\ref{eq:q5}) is impossible.
\par
\vspace{15pt} \centerline{\bf Case (c).} \vspace{10pt}\noindent If
vectors $\boldsymbol{\gamma}_2$ and $\boldsymbol{\gamma}_3$ are
linearly independent  no solution exists. Hence we may suppose
that vectors $\boldsymbol{\gamma}_2$ and $\boldsymbol{\gamma}_3$
are linearly dependent. Nevertheless, if
$\boldsymbol{\gamma}_3=\lambda\boldsymbol{\gamma}_2$ we proved
that eventual solutions lead to correlated properties, so we may
suppose the existence of a complex number $\lambda_4\neq\lambda$,
such that $\boldsymbol{\gamma}_3=\lambda_4\boldsymbol{\gamma}_2$.
As a consequence (iv-D) becomes
\begin{equation}\protect\label{eq:q6}
    \left\{\begin{array}{ll}
             \lbrack(q_{j1}\lambda_2+q_{j2})+\lambda_4(q_{j1}\lambda_3+q_{j3})\rbrack\boldsymbol{\gamma}_2=\boldsymbol{\gamma}_j\cr
             \lbrack(q_{j1}\lambda_2+q_{j2})+\lambda(q_{j1}\lambda_3+q_{j3})\rbrack\boldsymbol{\delta}_2={\bf 0}.
    \end{array}\right.
\end{equation}
Since  $\boldsymbol{\delta}_2\neq{\bf 0}$, second equation in
(\ref{eq:q6}) implies
$(q_{j1}\lambda_2+q_{j2})=-\lambda(q_{j1}\lambda_3+q_{j3})$, so
that first equation in (\ref{eq:q6}) yields
$(q_{j1}\lambda_3+q_{j3})(\lambda_4-\lambda)\boldsymbol{\gamma}_2=\boldsymbol{\gamma}_j$.\par
Straightforward calculations lead to a matrix $Q$ of the form
\begin{equation*}
Q={\left(\begin{array}{ccc}
       q_{11}&-\frac{\lambda(\lambda_2+\lambda_3\lambda_4)}{\lambda_4-\lambda}-q_{11}\lambda_2
       &\frac{\lambda_2+\lambda_3\lambda_4}{\lambda_4-\lambda}-q_{11}\lambda_3\cr\vspace{10pt}
       q_{21} & -\frac{\lambda}{\lambda_4-\lambda}-q_{21}\lambda_2 &\frac{1}{\lambda_4-\lambda}-q_{21}\lambda_3\cr\vspace{10pt}
       q_{31} &
       -\frac{\lambda\lambda_4}{\lambda_4-\lambda}-q_{31}\lambda_2 &\frac{\lambda_4}{\lambda_4-\lambda}-q_{31}\lambda_3\cr
   \end{array}\right).}
\end{equation*}
Nevertheless, self-adjointness of $Q$ yields to rather difficult
calculation.
\par
\vspace{15pt} \centerline{\bf Case (d).} \vspace{10pt}\noindent
Now we suppose $\boldsymbol{\gamma}_2={\bf 0}$ and
$\boldsymbol{\gamma}_3\neq {\bf 0}$ in order to make easier the
search of solutions. Equations in (iv-D) become
\begin{equation}\protect\label{eq:q8}
       \left\{\begin{array}{ll}
                    (q_{j1}\lambda_3+q_{j3})\boldsymbol{\gamma}_3=\boldsymbol{\gamma}_j\\
                    \lbrack(q_{j1}\lambda_2+q_{j2})+\lambda(q_{j1}\lambda_3+q_{j3})\rbrack\boldsymbol{\delta}_2={\bf 0}.
       \end{array}\right.
\end{equation}
Thus,  first equation of (\ref{eq:q8}) gets
\begin{equation*}
        \left\{\begin{array}{lll}
                     (q_{11}\lambda_3+q_{13})\boldsymbol{\gamma}_3=\lambda_3\boldsymbol{\gamma}_3\\
                     (q_{21}\lambda_3+q_{23})\boldsymbol{\gamma}_3={\bf 0}\\
                     (q_{31}\lambda_3+q_{33})\boldsymbol{\gamma}_3=\boldsymbol{\gamma}_3
        \end{array}\right.
\end{equation*}
and, since $\boldsymbol{\gamma}_3\neq 0$, this is equivalent to
say
\begin{equation*}
         \left\{\begin{array}{lll}
                          q_{13}=\lambda_3(1-q_{11})\\
                          q_{23}=-\lambda_3 q_{21}\\
                          q_{33}=1-\lambda_3 q_{31}
         \end{array}\right.
\end{equation*}
Similarly, second equation in (\ref{eq:q8}) implies
$$(q_{j1}\lambda_2+q_{j2})=-\lambda(q_{j1}\lambda_3+q_{j3}),$$
then
\begin{equation*}
     \left\{\begin{array}{lll}
                        q_{12}=-\lambda\lambda_3-\lambda_2 q_{11}\\
                        q_{22}=-\lambda_2 q_{21}\\
                        q_{32}=-\lambda-\lambda_2 q_{31}
           \end{array}\right.
\end{equation*}
By imposing self-adjointness, we find that $q_{11}=q$ is a real
number,
$\lambda=-\frac{\lambda_2\overline{\lambda}_3}{|{\lambda}_3|^2+1}$
and
\begin{equation*}
Q={\left(\begin{array}{ccc}
             q &-\lambda_2\left(q-\frac{{|\lambda_3|}^2}{1+{|\lambda_3|}^2}\right)
             &\lambda_3\left(1-q\right)\cr\\
             -\overline{\lambda}_2\left(q-\frac{{|\lambda_3|}^2}{1+{|\lambda_3|}^2}\right)
                    &|\lambda_2|^2 \left(q-\frac{{|\lambda_3|}^2}{1+{|\lambda_3|}^2}\right) &\lambda_3\overline{\lambda}_2
                    \left(q-\frac{{|\lambda_3|}^2}{1+{|\lambda_3|}^2}\right)\cr\\
             \overline{\lambda}_3\left(1-q\right) &\overline{\lambda}_3\lambda_2\left(q-\frac{{|\lambda_3|}^2}{1+{|\lambda_3|}^2}\right)
                   &1-{|\lambda_3|}^2\left(1-q\right)
 \end{array} \right).}
\end{equation*}
\par Taking into account
(\ref{eq:u2}), matrix $U$ has the form
\begin{equation*}
U={\left(
  \begin{array}{ccc}
             u_{11} &-u_{11}\lambda_2 &-u_{11}\lambda_3\cr
             u_{21} &-u_{21}\lambda_2 &-u_{21}\lambda_3\cr
             u_{11} &-u_{31}\lambda_2 &-u_{31}\lambda_3\cr
  \end{array}
  \right).}
\end{equation*}
Analogous reasonings, for systems (i-A) and (iii-C), provides
matrices  $P$ and $V$ of a similar form; however, self-adjointness
of $G$ implies that $V=\overline{U}^t$, in such a manner that we
attain the solution presented in  section 2.2.

\section{Detection of four incompatible properties}
We already noticed the difference between the meaning of our results with respect to that obtained by VAA; moreover, they affirm \cite{[Vaid]} that, according to their method, it is no possible to produce inferences (such that described in \cite{[Vaid]}) for more than three observable. Since our method runs in a quite different matter, maybe it shall admit solutions.\\
In this section we treat the question whether two incompatible properties, $G$ and $L$, can be detected together with the further incompatible property $E$ (Wich-Slit property)  and together with measurement of  the final impact point,
in the same kind of ideal experiment.\\
More precisely, we seek for a concrete Hilbert space ${\cal H}={\cal H}_I\otimes{\cal H}_{II}$ where $E$ is the projection operator acting on ${\cal H}_I$ representing which-slit property, such that a concrete state $\Psi$, and concrete projection operators $G$ and $L$  representing mutually incompatible properties can be found in such a manner that:
\begin{itemize}
\item[-] Property $E$ can be detected by means of a detector $T$, acting on ${\cal H}_{II}$
\item[-] Property $G$ can be detected by means of a detector $Y$, acting on ${\cal H}_{II}$
\item[-] Property $L$ can be detected by means of a detector $W$, acting on ${\cal H}_{II}$
\item[-] The three detections can be carried out togheter, i.e. $[T,Y]=0$, $[T,W]=0$, $[Y,W]=0$.
\end{itemize}
The method used for such a research is similar to that presented in previous section.\\
In the rest of this section we formulate the question in formal way as problem \emph{(P)}; as before, we adopt a matrix representation and in this framework we establish the constraints to be satisfied in order that  solutions exist. Then we present concrete solutions.\\
The systematic research of solution is in suitable section 6. We set out such a research in order to answer the question whether non-correlated solutions do exist or not.
Hence, the case $dim({\cal H}_I)=10$ is investigated; a detailed analysis show that solutions exist, however, the three detections turn out to be always correlated.

\subsection{\bf Mathematical Formalism}
Let  $G=G_{I}\otimes{\bf 1}_{II}$ and $L=L_{I}\otimes{\bf 1}_{II}$ be properties incompatible with WS property $E$. Detection of both $G$, $L$ and   which-slit property $E$ is possible if, with respect to the same state vector $\Psi$, there exists a which-slit  detector $T={\bf 1}_{I}\otimes T_{II}$ of $E$, a  detector $Y={\bf 1}_{I}\otimes Y_{II}$ of $G$ and a detector $W={\bf 1}_{I}\otimes W_{II}$ of $L$. Let detectors satisfy the condition $[T,Y]=[T,W]=[Y, W]=0$ in such a manner that $T$, $Y$ and $W$ can be measured together; hence they provide  simultaneous informations about  $E$, $G$ and $L$. \\
Formally, we are asking if the following problem has solution
\\[10pt]
\emph{(P) Given the property $E=E_{I}\otimes {\bf 1}_{II}$ we have to find
\begin{itemize}
\item[-]  projection operators $G_{I}$ and $L_{I}$ of ${\cal H}_I$
\item[-]  projection operators $T_{II}$, $Y_{II}$ and $W_{II}$ of ${\cal H}_{II}$
\item[-] a state vector $\Psi\in {\cal H}_I\otimes {\cal H}_{II}$
\end{itemize}
such that the following conditions are satisfied:
\begin{enumerate}
\item[(C.1)] $[E,G]\neq 0$ i.e $[E_{I},G_{I}]\neq 0$%1
\item[(C.2)] $[E,L]\neq 0$ i.e $[E_{I},L_{I}]\neq 0$%2
\item[(C.3)] $[L,G]\neq 0$ i.e $[L_{I},G_{I}]\neq 0$%3
\item[(C.4)] $[T,E]= 0$ and $T\Psi= E\Psi$ %4
\item[(C.5)] $[Y,G]= 0$ and $Y\Psi= G\Psi$ %5
\item[(C.6)] $[W,L]= 0$ and $W\Psi= L\Psi$ %6
\item[(C.7)] $[T,Y]= 0$ %7
\item[(C.8)] $[T,W]= 0$ %8
\item[(C.9)] $[Y,W]=0$ %9
\item[(C.10)] $\Psi\neq E\Psi\neq 0$, $\Psi\neq G\Psi\neq 0$ and $\Psi\neq L\Psi\neq 0$ %10
\end{enumerate}}
\noindent
Conditions (C.1)-(C.3) are equivalent to state that properties represented by projection operators $E$, $G$ and $L$ are mutually non-compatible. In the remaining items, the fact that the commutators are zero
 is expression of the compatibility (hence simultaneous knowledge) between properties represented by  the projection operators involved.  Moreover, for a given state $\Psi$ if equation   $T\Psi= E\Psi$ in (C.4)
(resp. $Y\Psi= G\Psi$ in (C.5) and $W\Psi= L\Psi$ in (C.6)) holds, then it is also possible to detect which slit each particle passes through (resp. to detect $G$ or $I-G$, to detect $L$ or $I-L$), by means of $T$ (resp. $Y$ and $W$) ; indeed, the formula
$$p(T\mid E)=\frac{\langle\Psi\mid TE\Psi\rangle}{\langle\Psi\mid E\Psi\rangle}=1=\frac{\langle\Psi\mid TE\Psi\rangle}{\langle\Psi\mid T\Psi\rangle}=p(E\mid T)
$$
represents the conditional probabilities allowing us to infer the passage of particle 1 through slit 1 from the occurrence  of outcome 1 for $T$: in this sense $E$ and $T$ are correlated properties (similarly in (C.5) and (C.6)).
Condition (C.10) is added to exclude solutions corresponding to uninteresting case that $\Psi$ is eigenvector of $E$, $G$ or $L$.\\
We introduce matrix representation, to make easier our task.

\subsection{\bf Matrix Representation}
${\cal H}_I$ representation decribed in section II is general so that it can be adopted here without modifications.
Therefore, projection operators $E_{I}$,$G_{I}$ and $L_{I}$ in (C.1)-(C.10) must have the following representations:
\begin{equation}\protect\label{eq:I_representation}
E_{I}={\left(
\begin{array}{cc}
 {\bf 1} & {\bf 0} \\
 {\bf 0} & {\bf 0}
\end{array}
\right)},
\quad
G_{I}={\left(
\begin{array}{cc}
 P & U \\
 V & Q
\end{array}
\right)},
\quad
L_{I}={\left(
\begin{array}{cc}
 M & Z \\
 W & N
\end{array}
\right)}
\end{equation}
where $U\neq {\bf 0}$ and $Z\neq {\bf 0}$; $G_{I}={G_{I}}^{\ast}={G_{I}}^2$ and $L_{I}={L_{I}}^{\ast}={L_{I}}^2$.
Constraints $U\neq{\bf 0}$ and $Z\neq {\bf 0}$ above are equivalent to $[E_{I},G_{I}]\neq {\bf 0}$ and $[E_{I},L_{I}]\neq {\bf 0}$ required by (C.1) and (C.2).\\
%{\bf ${\cal H}_{II}$ Representation. }
Following the same argument of previous section for the representation of ${\cal H}_{II}$, eight projection operators, $A_i$ ($i=1,\ldots,8$) of ${\cal H}_{II}$  must exist, such that $\sum_1^8 A_i={\bf 1}$, $T_{II}=A_1+A_2+A_3+A_5$, $Y_{II}=A_1+A_2+A_4+A_6$, $W_{II}=A_1+A_3+A_4+A_7$ (fig.2).
\begin{figure}[h]\label{figura2}
\begin{center}
\includegraphics[height=4cm]{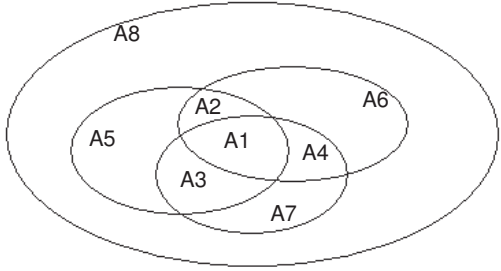}
\caption{representation for ${\cal H}_{II}$}
\end{center}
\end{figure}
\noindent
Then we choose to represent every vector  ${\bf x}\in {\cal H}_{II}$ as a column vector
${\bf x}=({\bf a},{\bf b}, {\bf c}, {\bf d}, {\bf e}, {\bf f}, {\bf g}, {\bf h})^t$
where
${\bf a}=A_1{\bf x}, {\bf b}=A_2{\bf x},\ldots, {\bf h}=A_8{\bf x}$.
As a consequence, projection operators $T_{II}$, $Y_{II}$, $W_{II}$ in (C.1)-(C.10) must satisfy the following constraints
$$
T_{II}=
{\left(
\begin{array}{cccccccc}
 {\bf 1} & {\bf 0} & {\bf  0} & {\bf 0} & {\bf 0} & {\bf 0} & {\bf 0} & {\bf 0} \\
 {\bf 0} & {\bf 1} & {\bf 0} & {\bf 0} & {\bf 0} & {\bf 0} & {\bf 0} & {\bf 0} \\
 {\bf 0} & {\bf 0} & {\bf 1} & {\bf 0} & {\bf 0} & {\bf 0} & {\bf 0} & {\bf 0} \\
 {\bf 0} & {\bf 0} & {\bf 0} & {\bf 0} & {\bf 0} & {\bf 0} & {\bf 0} & {\bf 0} \\
 {\bf 0} & {\bf 0} & {\bf 0} & {\bf 0} & {\bf 1} & {\bf 0} & {\bf 0} & {\bf 0} \\
 {\bf 0} & {\bf 0} & {\bf 0} & {\bf 0} & {\bf 0} & {\bf 0} & {\bf 0} & {\bf 0} \\
 {\bf 0} & {\bf 0} & {\bf 0} & {\bf 0} & {\bf 0} & {\bf 0} & {\bf 0} & {\bf 0} \\
 {\bf 0} & {\bf 0} & {\bf 0} & {\bf 0} & {\bf 0} & {\bf 0} & {\bf 0} & {\bf 0}
\end{array}
\right)},\qquad
Y_{II}=
{\left(
\begin{array}{cccccccc}
 {\bf 1} & {\bf 0} & {\bf  0} & {\bf 0} & {\bf 0} & {\bf 0} & {\bf 0} & {\bf 0} \\
 {\bf 0} & {\bf 1} & {\bf 0} & {\bf 0} & {\bf 0} & {\bf 0} & {\bf 0} & {\bf 0} \\
 {\bf 0} & {\bf 0} & {\bf 0} & {\bf 0} & {\bf 0} & {\bf 0} & {\bf 0} & {\bf 0} \\
 {\bf 0} & {\bf 0} & {\bf 0} & {\bf 1} & {\bf 0} & {\bf 0} & {\bf 0} & {\bf 0} \\
 {\bf 0} & {\bf 0} & {\bf 0} & {\bf 0} & {\bf 0} & {\bf 0} & {\bf 0} & {\bf 0} \\
 {\bf 0} & {\bf 0} & {\bf 0} & {\bf 0} & {\bf 0} & {\bf 1} & {\bf 0} & {\bf 0} \\
 {\bf 0} & {\bf 0} & {\bf 0} & {\bf 0} & {\bf 0} & {\bf 0} & {\bf 0} & {\bf 0} \\
 {\bf 0} & {\bf 0} & {\bf 0} & {\bf 0} & {\bf 0} & {\bf 0} & {\bf 0} & {\bf 0}
\end{array}
\right)}
$$
and
\begin{equation}\protect\label{eq:II_representation}
W_{II}=
{\left(
\begin{array}{cccccccc}
 {\bf 1} & {\bf 0} & {\bf  0} & {\bf 0} & {\bf 0} & {\bf 0} & {\bf 0} & {\bf 0} \\
 {\bf 0} & {\bf 0} & {\bf 0} & {\bf 0} & {\bf 0} & {\bf 0} & {\bf 0} & {\bf 0} \\
 {\bf 0} & {\bf 0} & {\bf 1} & {\bf 0} & {\bf 0} & {\bf 0} & {\bf 0} & {\bf 0} \\
 {\bf 0} & {\bf 0} & {\bf 0} & {\bf 1} & {\bf 0} & {\bf 0} & {\bf 0} & {\bf 0} \\
 {\bf 0} & {\bf 0} & {\bf 0} & {\bf 0} & {\bf 0} & {\bf 0} & {\bf 0} & {\bf 0} \\
 {\bf 0} & {\bf 0} & {\bf 0} & {\bf 0} & {\bf 0} & {\bf 0} & {\bf 0} & {\bf 0} \\
 {\bf 0} & {\bf 0} & {\bf 0} & {\bf 0} & {\bf 0} & {\bf 0} & {\bf 1} & {\bf 0} \\
 {\bf 0} & {\bf 0} & {\bf 0} & {\bf 0} & {\bf 0} & {\bf 0} & {\bf 0} & {\bf 0}
\end{array}
\right)}
\end{equation}
%{\bf ${\cal H}_{I}\otimes{\cal H}_{II} $ representation. }
The ${\cal H}={\cal H}_{I}\otimes{\cal H}_{II} $ representation described in section II can be adopted here taking into account that
every vector $\Psi$ in the product space ${\cal H}_I\otimes {\cal H}_{II}$ shall be represented as a column vector
$$\Psi=^t\left({\bf x}_1, {\bf x}_2, \ldots,{\bf x}_i,\ldots;{\bf y}_1, {\bf y}_2,\ldots,{\bf y}_j,\ldots\right)$$
where
$
{\bf x}_i=^t\left({\bf a}_i, {\bf b}_i, {\bf c}_i, {\bf d}_i, {\bf e}_i, {\bf f}_i, {\bf g}_i, {\bf h}_i\right)$ and
${\bf y}_j=^t\left(\boldsymbol{\alpha}_j,\boldsymbol{\beta}_j,\boldsymbol{\gamma}_j, \boldsymbol{\delta}_j,\boldsymbol{\epsilon}_j,\boldsymbol{\zeta}_j,\boldsymbol{\eta}_j,\boldsymbol{\theta}_j\right)
$
and
${\bf a}_i=A_1{\bf x}_i, {\bf b}_i=A_2{\bf x}_i,\ldots,
{\bf h}_i=A_8{\bf x}_i,
\boldsymbol{\alpha}_j=A_1{\bf y}_j, \boldsymbol{\beta}_j=A_2{\bf y}_j,\ldots,
\boldsymbol{\theta}_j=A_8{\bf y}_j$.

\subsection{\bf Constraints for $\Psi$, $G_{I}$, $L_{I}$}
According to (\protect\ref{eq:otimes}), (\protect\ref{eq:I_representation}) and  (\protect\ref{eq:II_representation})
$E=E_{I}\otimes {\bf 1}$ and $T={\bf 1}\otimes T_{II}$ are represented in matrix form as
\begin{equation}
E=
{\left(
\begin{array}{cccccccc}
 {\bf 1} & {\bf 0} & \ldots & {\bf 0} & {\bf 0} & \ldots \\
 {\bf 0} & {\bf 1} & \ldots & {\bf 0} & {\bf 0} & \ldots   \\
 \vdots & \vdots & \ddots & \vdots & \vdots & \ddots  \\
 {\bf 0} & {\bf 0} & \ldots & {\bf 0} & {\bf 0} & \ldots  \\
 {\bf 0} & {\bf 0} & \ldots & {\bf 0} & {\bf 0} & \ldots \\
 \vdots & \vdots & \ddots & \vdots & \vdots & \ddots  \\
 \end{array}
\right)}
\quad
T=
{\left(
\begin{array}{cccccc}
 {\bf T}_{\bf {II}} & {\bf 0} & \ldots & {\bf 0} & {\bf 0} & \ldots \\
 {\bf 0} & {\bf T}_{\bf {II}} & \ldots & {\bf 0} & {\bf 0} & \ldots   \\
 \vdots & \vdots & \ddots & \vdots & \vdots & \ddots  \\
 {\bf 0} & {\bf 0} & \ldots & {\bf T}_{\bf {II}} & {\bf 0} & \ldots  \\
 {\bf 0} & {\bf 0} & \ldots & {\bf 0} & {\bf T}_{\bf {II}} & \ldots \\
 \vdots & \vdots & \ddots & \vdots & \vdots & \ddots  \\
\end{array}
\right)}\end{equation}
Condition (C.4) $T\Psi=E\Psi$ implies ${\bf d}_{ i}={\bf f}_{i}={\bf g}_i={\bf h}_i={\bf 0}$ and $\boldsymbol{\alpha}_j=\boldsymbol{\beta}_j=\boldsymbol{\gamma}_j=\boldsymbol{\epsilon}_j={\bf 0}$ so that ${\bf x}_i$ and ${\bf y}_j$ take the form
\begin{equation}\protect\label{eq:first constraint }
\begin{array}{ll}
{\bf x}_i=\left({\bf a}_i, {\bf b}_i, {\bf c}_i, {\bf 0}, {\bf e}_i, {\bf 0}, {\bf 0}, {\bf 0}\right)\\[10pt]
{\bf y}_j=\left({\bf 0},  {\bf 0},  {\bf 0}, \boldsymbol{\delta}_j, {\bf 0}, \boldsymbol{\zeta}_j, \boldsymbol{\eta}_j, \boldsymbol{\theta}_j\right)\end{array}
\end{equation}
%{\bf Constraints imposed by $Y\Psi=G\Psi$}
Further constraints are imposed by condition (C.5)  $Y\Psi=G\Psi$.
Since $G_{I}={\left(
\begin{array}{cccc}
 P & U \\
 V & Q
\end{array}
\right)}$ then
projection operators $G=G_{I}\otimes {\bf 1}$ and $Y={\bf 1}\otimes Y_{II}$ are represented as
\begin{equation}\protect\label{eq:GYtensor}
G={\left(
\begin{array}{cccccc}
 p_{11}{\bf 1} & p_{12}{\bf 1} & \ldots & u_{11}{\bf 1} & u_{12}{\bf 1} & \ldots \\
 p_{21}{\bf 1} & p_{22}{\bf 1} & \ldots & u_{21}{\bf 1} & u_{22}{\bf 1} & \ldots   \\
 \vdots & \vdots & \ddots & \vdots & \vdots & \ddots  \\
 v_{11}{\bf 1} & v_{12}{\bf 1} & \ldots & q_{11}{\bf 1} & q_{12}{\bf 1} & \ldots  \\
 v_{21}{\bf 1} & v_{22}{\bf 1} & \ldots & q_{21}{\bf 1} & q_{22}{\bf 1} & \ldots \\
 \vdots & \vdots & \ddots & \vdots & \vdots & \ddots  \\
\end{array}
\right)}
\quad
Y=
{\left(
\begin{array}{cccccc}
 {\bf Y}_{\bf {II}} & {\bf 0} & \ldots & {\bf 0} & {\bf 0} & \ldots \\
 {\bf 0} & {\bf Y}_{\bf {II}} & \ldots & {\bf 0} & {\bf 0} & \ldots   \\
 \vdots & \vdots & \ddots & \vdots & \vdots & \ddots  \\
 {\bf 0} & {\bf 0} & \ldots & {\bf Y}_{\bf {II}} & {\bf 0} & \ldots  \\
 {\bf 0} & {\bf 0} & \ldots & {\bf 0} & {\bf Y}_{\bf {II}} & \ldots \\
 \vdots & \vdots & \ddots & \vdots & \vdots & \ddots  \\
\end{array}
\right)}
\end{equation}
and
\begin{equation}\protect\label{eq:GPsi}\begin{array}{ll}
Y\Psi=^t (\cdots, {\bf a}_i, {\bf b}_i, {\bf 0}, {\bf 0}, {\bf 0}, {\bf 0}, {\bf 0}, {\bf 0}; {\bf 0}, {\bf 0}, {\bf 0},\boldsymbol{\delta}_j, {\bf 0}, \boldsymbol{\zeta}_j, {\bf 0}, {\bf 0},\cdots)\\
G\Psi=^t ({\bf z}_1,{\bf z}_2,\cdots,{\bf z}_i,\cdots;{\bf w}_1,{\bf w}_2,\cdots,{\bf w}_j,\cdots)
\end{array}
\end{equation}\protect\label{eq:ypsi}
where
\begin{equation}\protect\label{eq:z_and_v}
{\bf z}_i=
{\left(
\begin{array}{c}
 \sum_k p_{ik}\bf{a}_k \\
\sum_k p_{ik}\bf{b}_k \\
\sum_k p_{ik}\bf{c}_k \\
\sum_k u_{ik}\boldsymbol{\delta}_k \\
\sum_k p_{ik}\bf{e}_k \\
\sum_k u_{ik}\boldsymbol{\zeta}_k \\
\sum_k u_{ik}\boldsymbol{\eta}_k \\
\sum_k u_{ik}\boldsymbol{\theta}_k \\
\end{array}
\right)}\qquad
\textrm {and}\qquad
{\bf w}_j=
{\left(
\begin{array}{c}
 \sum_k v_{jk}{\bf  a}_k \\
\sum_k v_{jk}{\bf b}_k \\
\sum_k v_{jk}{\bf c}_k \\
\sum_k q_{jk}\boldsymbol{\delta}_k \\
\sum_k v_{jk}{\bf e}_k \\
\sum_k q_{jk} \boldsymbol{\zeta}_k \\
\sum_k q_{jk}\boldsymbol{\eta}_k \\
\sum_k q_{jk}\boldsymbol{\theta}_k \\
\end{array}
\right)}
\end{equation}
so that, taking into account (\protect\ref{eq:GYtensor}) and (\protect\ref{eq:z_and_v}), condition (C.5) $Y\Psi=G\Psi$ can be explicited as
\begin{equation}\protect\label{eq:sistema I}
\begin{array}{ll}
(i-A)&\quad
\left\{ \begin{array}{ll}
\sum_k p_{ik}{\bf a}_k={\bf a}_i \\
\sum_k p_{ik}{\bf b}_k={\bf b}_i\\
\sum_k p_{ik}{\bf c}_k={\bf 0} \\
\sum_k p_{ik}{\bf e}_k={\bf 0}
\end{array} \right.
\qquad
(ii-B)\quad
\left\{ \begin{array}{ll}
\sum_k u_{ik}\boldsymbol{\delta}_k={\bf 0} \\
\sum_k u_{ik} \boldsymbol{\zeta}_k={\bf 0} \\
\sum_k u_{ik}\boldsymbol{\eta}_k={\bf 0} \\
\sum_k u_{ik}\boldsymbol{\theta}_k={\bf 0} \\
\end{array} \right.\\
\\[1pt]
(iii-C)&\quad
\left\{ \begin{array}{ll}
\sum_k v_{ik}{\bf a}_k={\bf 0} \\
\sum_k v_{ik}{\bf b}_k={\bf 0} \\
\sum_k v_{ik}{\bf c}_k={\bf 0} \\
\sum_k v_{ik}{\bf e}_k={\bf 0}
\end{array} \right.
\qquad
(iv-D)\quad
\left\{ \begin{array}{ll}
\sum_k q_{ik}\boldsymbol{\delta}_k=\boldsymbol{\delta}_i \\
\sum_k q_{ik} \boldsymbol{\zeta}_k=\boldsymbol{\zeta}_i \\
\sum_k q_{ik}\boldsymbol{\eta}_k={\bf 0} \\
\sum_k q_{ik}\boldsymbol{\theta}_k={\bf 0}
\end{array} \right.
\end{array}
\end{equation}
%\subsection{Constraints imposed by $W\Psi=L\Psi$}
Further constraints are imposed by condition (C.6) $Y\Psi=G\Psi$.
Since $L_{I}={\left(
\begin{array}{cccc}
 M & Z \\
 W & N
\end{array}
\right)}$ then,
%according to (\protect\ref{eq:tensor}) and (\protect\ref{eq:first constraint }),
projection operators $L=L_{I}\otimes {\bf 1}$ and $W={\bf 1}\otimes W_{II}$ are represented as
\begin{equation}
L={\left(
\begin{array}{cccccc}
 m_{11}{\bf 1} & m_{12}{\bf 1} & \ldots & z_{11}{\bf 1} & z_{12}{\bf 1} & \ldots \\
 m_{21}{\bf 1} & m_{22}{\bf 1} & \ldots & z_{21}{\bf 1} & z_{22}{\bf 1} & \ldots   \\
 \vdots & \vdots & \ddots & \vdots & \vdots & \ddots  \\
 w_{11}{\bf 1} & w_{12}{\bf 1} & \ldots & n_{11}{\bf 1} & n_{12}{\bf 1} & \ldots  \\
 w_{21}{\bf 1} & w_{22}{\bf 1} & \ldots & n_{21}{\bf 1} & n_{22}{\bf 1} & \ldots \\
 \vdots & \vdots & \ddots & \vdots & \vdots & \ddots  \\
\end{array}
\right)}
\quad
W=
{\left(
\begin{array}{cccccc}
 {\bf W_{II}} & {\bf 0} & \ldots & {\bf 0} & {\bf 0} & \ldots \\
 {\bf 0} & {\bf W_{II}} & \ldots & {\bf 0} & {\bf 0} & \ldots   \\
 \vdots & \vdots & \ddots & \vdots & \vdots & \ddots  \\
 {\bf 0} & {\bf 0} & \ldots & {\bf W_{II}} & {\bf 0} & \ldots  \\
 {\bf 0} & {\bf 0} & \ldots & {\bf 0} & {\bf W_{II}} & \ldots \\
 \vdots & \vdots & \ddots & \vdots & \vdots & \ddots  \\
\end{array}
\right)}
\end{equation}
and
\begin{equation}\protect\label{eq:LPsi}
\begin{array}{ll}
W\Psi=^t(\cdots, {\bf a}_i,{\bf 0},{\bf c}_i,{\bf 0},{\bf 0},{\bf 0},{\bf 0},{\bf 0};{\bf 0},{\bf 0},{\bf 0},\boldsymbol{\delta}_j,{\bf 0},{\bf 0},\boldsymbol{\eta}_j,0,\cdots)\\
L\Psi=^t({\bf s}_1,{\bf s}_2,\cdots,{\bf s}_i,\cdots;{\bf t}_1,{\bf t}_2,\cdots,{\bf t}_j,\cdots)
\end{array}
\end{equation}
where
\begin{equation}
{\bf s}_i=
{\left(
\begin{array}{c}
 \sum_k m_{ik}{\bf a}_k \\
\sum_k m_{ik}{\bf b}_k \\
\sum_k m_{ik}{\bf c}_k \\
\sum_k z_{ik}\boldsymbol{\delta}_k \\
\sum_k m_{ik}{\bf e}_k \\
\sum_k z_{ik} \boldsymbol{\zeta}_k \\
\sum_k z_{ik}\boldsymbol{\eta}_k \\
\sum_k z_{ik}\boldsymbol{\theta}_k \\
\end{array}
\right)}\qquad
\textrm {and}\qquad
{\bf t_j}=
{\left(
\begin{array}{c}
 \sum_k w_{jk}{\bf a}_k \\
\sum_k w_{jk}{\bf b}_k \\
\sum_k w_{jk}{\bf c}_k \\
\sum_k n_{jk}\boldsymbol{\delta}_k \\
\sum_k w_{jk}{\bf e}_k \\
\sum_k n_{jk} \boldsymbol{\zeta}_k \\
\sum_k n_{jk}\boldsymbol{\eta}_k \\
\sum_k n_{jk}\boldsymbol{\theta}_k \\
\end{array}
\right)}.
\end{equation}
Then condition (C.6) $W\Psi=L\Psi$, can be written as
\begin{equation}\protect\label{eq:sistema II}
\begin{array}{ll}
(i-A')\quad
&\left\{ \begin{array}{ll}
\sum_k m_{ik}{\bf a}_k={\bf a}_i \\
\sum_k m_{ik}{\bf b}_k={\bf 0}\\
\sum_k m_{ik}{\bf c}_k={\bf c}_i \\
\sum_k m_{ik}{\bf e}_k={\bf 0}
\end{array} \right.
\qquad
(ii-B')\quad
\left\{ \begin{array}{ll}
\sum_k z_{ik}\boldsymbol{\delta}_k={\bf 0} \\
\sum_k z_{ik} \boldsymbol{\zeta}_k={\bf 0} \\
\sum_k z_{ik}\boldsymbol{\eta}_k={\bf 0}\\
\sum_k z_{ik}\boldsymbol{\theta}_k={\bf 0} \\
\end{array} \right.\\
\\[1pt]
(iii-C')\quad
&\left\{ \begin{array}{ll}
\sum_k w_{ik}{\bf a}_k={\bf 0} \\
\sum_k w_{ik}{\bf b}_k={\bf 0} \\
\sum_k w_{ik}{\bf c}_k={\bf 0} \\
\sum_k w_{ik}{\bf e}_k={\bf 0}
\end{array} \right.
\qquad
(iv-D')\quad
\left\{ \begin{array}{ll}
\sum_k n_{ik}\boldsymbol{\delta}_k=\boldsymbol{\delta}_i \\
\sum_k n_{ik} \boldsymbol{\zeta}_k={\bf 0}\\
\sum_k n_{ik}\boldsymbol{\eta}_k=\boldsymbol{\eta}_j \\
\sum_k n_{ik}\boldsymbol{\theta}_k={\bf 0}
\end{array} \right.
\end{array}
\end{equation}

\section{A family of solutions}
Until now
we have established general constraints to be satisfied by any solution of the problem, independently of the ranks of matrices, and then of dimensions of the spaces ${\cal H}_{I}$ and ${\cal H}_{II}$. Here we present a concrete solution of the problem, whose derivation is displaced in next section; our research is not at all exhaustive: we analyze a particular situation and, according to it, the three detections turn out to be always correlated.\\
We notice that, if in correspondence of a given state $\Psi$ satisfying (\protect\ref{eq:first constraint }), matrices $G$ and $L$, solutions of (\protect\ref{eq:sistema I}) and (\protect\ref{eq:sistema II}) exist such that $[E,G]$, $[G,L]$ and $[L,E]$ are non zero, $G=G^{\star}=G^2$ and   $L=L^{\star}=L^2$, then, all (C-4)-(C-9) are automatically satisfied.\\
As in previous case, we restrict our search to the case that the two slit are symmetrical: this leads to exclude odd dimensions of ${\cal H}_{I}$ and, moreover, to assume that $rank(L)=rank(I-L)=dim({\cal H}_{I})/2$.\\
We shall proceed as follows: at the beginning, no constrain is imposed about the dimension of ${\cal H}_1$. We restrict our search by working with subsystems (ii-B) and (iv-D) of (\protect\ref{eq:sistema I}) and (ii-B') and (iv-D') of (\protect\ref{eq:sistema II}), rather than with the entire systems (\protect\ref{eq:sistema I}) and (\protect\ref{eq:sistema II}); then we take analogous results for the remaining subsystems (which have analogous forms).
By using the elementary notion of linear combination, we make some hypothesis of linear dependence or independence among the vector-components  of ${\bf x}_i$ and  ${\bf y}_j$; since no hypothesis is made about the dimension of ${\cal H}_I$, we can suppose $i$ running from 1 to n.\\
Our task would be more meaningful if we attain non-correlated solutions of \emph(P); for such a  reason, at this level of the search, we neglect the correlated ones we gradually find. So, we reduce systems (\protect\ref{eq:sistema I}) and (\protect\ref{eq:sistema II}) in a more useful form, which  makes evident that, with our assumptions, if solutions exist, they are always correlated. Thereby, we shall see that concrete solutions of (C-1)-(C-10) exist; at this point, our task is made easier by fixing the dimension of ${\cal H}_{I}$, $dim({\cal H}_{I})=10$, and searching solutions corresponding to a particular state vector $\Psi$(see (\protect\ref{eq:choice})). Taking into account  self-adjointness of $G_{I}={\left(
\begin{array}{cc}
 P & U \\
 V & Q
\end{array}
\right)}$ and
$L_{I}={\left(
\begin{array}{cc}
 M & Z \\
 W & N
\end{array}
\right)}$ we get {\footnotesize
$$
Q=\left(\begin{array}{ccccc} q &
-\alpha_2\left(q-\frac{1}{\Gamma}\right) & -\alpha_3q &
-\beta_4\frac{\alpha_2}{\Gamma} & -\beta_5\frac{\alpha_2}{\Gamma}\\[10pt]
 -\overline{\alpha}_2\left(q-\frac{1}{\Gamma}\right) & \Lambda_3+
 {\mid \alpha_2\mid}^2\left(q-\frac{1}{\Gamma}\right) & \alpha_3
 \overline{\alpha}_2\left(q-\frac{1}{\Gamma}\right)& -\beta_4\Lambda_3 & -\beta_5\Lambda_3\\[10pt]
-\overline{\alpha}_3q &
-\overline{\alpha}_3\alpha_2\left(q-\frac{1}{\Gamma}\right) &
{\mid \alpha_3\mid}^2q &\overline{\alpha}_3\beta_4\frac{\alpha_2}{\Gamma} & \overline{\alpha}_3\beta_5\frac{\alpha_2}{\Gamma}\\[10pt]
-\overline{\beta}_4\frac{\overline{\alpha}_2}{\Gamma} &
-\overline{\beta}_4\Lambda_3 &
\overline{\beta}_4\alpha_3\frac{\overline{\alpha}_2}{\Gamma} & {\mid \beta_4\mid}^2\Lambda & \overline{\beta}_4\beta_5\Lambda\\[10pt]
-\overline{\beta}_5\frac{\overline{\alpha}_2}{\Gamma} &
-\overline{\beta}_5\Lambda_3 &
\overline{\beta}_5\alpha_3\frac{\overline{\alpha}_2}{\Gamma} &
\overline{\beta}_5\beta_4\Lambda & {\mid \beta_5\mid}^2\Lambda
\end{array}\right),
$$}
where
\begin{itemize}
   \item[-]$\Gamma=\left(1+{\mid\alpha_3\mid}^2\right)+\left({\mid \beta_4\mid}^2+{\mid
           \beta_5\mid}^2\right)\left(1+{\mid \alpha_2\mid}^2+{\mid
           \alpha_3\mid}^2\right)$,
   \item[-]$\Lambda_2=\frac{{\mid
           \alpha_2\mid}^2}{\Gamma}$,
   $\Lambda_3=\frac{1+{\mid
           \alpha_3\mid}^2}{\Gamma}$ and
   $\Lambda=\Lambda_2+\Lambda_3$;
   \end{itemize}
{\scriptsize{\setlength\arraycolsep{1.3pt}
$$ N=\left(\begin{array}{ccccc}
n & -\alpha_2\left(n-\frac{1}{\Gamma}\right) &
-\alpha_3\left(n-\frac{1}{\Delta}\right) &
-\beta_4\frac{\alpha_2}{\Gamma}-\lambda_4\frac{\alpha_3}{\Delta} & -\beta_5\frac{\alpha_2}{\Gamma}-\lambda_5\frac{\alpha_3}{\Delta}\\[10pt]
-\overline{\alpha}_2\left(n-\frac{1}{\Gamma}\right) & \Delta_3+
{\mid \alpha_2\mid}^2\left(n-\frac{1}{\Gamma}\right) & \alpha_3
\overline{\alpha}_2\left(n-\frac{1}{\Gamma}-
\frac{1}{\Delta}\right)&
-\beta_4\Delta_3+\lambda_4\overline{\alpha}_2\frac{\alpha_3}{\Delta}
&
-\beta_5\Delta_3+\lambda_5\overline{\alpha}_2\frac{\alpha_3}{\Delta}\\[10pt]
-\overline{\alpha}_3\left(n-\frac{1}{\Delta}\right) & \alpha_2
\overline{\alpha}_3\left(n-\frac{1}{\Gamma}-\frac{1}{\Delta}\right)
& \Sigma_2+{\mid \alpha_3\mid}^2\left(n-\frac{1}{\Delta}\right)
&\overline{\alpha}_3\beta_4\frac{\alpha_2}{\Gamma}-\lambda_4\Sigma_2
&
\overline{\alpha}_3\beta_5\frac{\alpha_2}{\Gamma}-\lambda_5\Sigma_2\\[10pt]
-\overline{\beta}_4\frac{\overline{\alpha}_2}{\Gamma}-\overline{\lambda}_4\frac{\overline{\alpha}_3}{\Delta}
&
-\overline{\beta}_4\Delta_3+\alpha_2\overline{\lambda}_4\frac{\overline{\alpha}_3}{\Delta}
&
\overline{\beta}_4\alpha_3\frac{\overline{\alpha}_2}{\Gamma}-\overline{\lambda}_4\Sigma_2
&
{\mid \beta_4\mid}^2\Delta+{\mid \lambda_4\mid}^2\Sigma & \overline{\beta}_4\beta_5\Delta+\overline{\lambda}_4\lambda_5\Sigma\\[10pt]
-\overline{\beta}_5\frac{\overline{\alpha}_2}{\Gamma}-\overline{\beta}_4\frac{\overline{\alpha}_3}{\Delta}
&
-\overline{\beta}_5\Delta_3+\alpha_2\overline{\beta}_4\frac{\overline{\alpha}_3}{\Delta}
&
\overline{\beta}_5\alpha_3\frac{\overline{\alpha}_2}{\Gamma}-\overline{\lambda}_5\Sigma_2
&
\overline{\beta}_5\beta_4\Delta+\overline{\lambda}_5\lambda_4\Sigma
& {\mid \beta_5\mid}^2\Delta+{\mid \lambda_5\mid}^2\Sigma
\end{array}\right),
$$}}
\noindent where
\begin{itemize}
      \item[-]$\lambda_4=\frac{\alpha_2\overline{\alpha}_3}{\overline{\beta}_4\left(1+{\mid
              \alpha_2\mid}^2+{\mid
              \alpha_3\mid}^2\right)}-\lambda_5\frac{\overline{\beta}_5}{\overline{\beta}_4}$,
      \item[-]$\Delta=\left(1+{\mid \alpha_2\mid}^2\right)+\left({\mid
              \lambda_4\mid}^2+{\mid \lambda_5\mid}^2\right)\left(1+{\mid
              \alpha_2\mid}^2+{\mid \alpha_3\mid}^2\right)$,
      \item[-]$\Sigma_2=\frac{1+{\mid \alpha_2\mid}^2}{\Delta}$,
              $\Sigma_3=\frac{{\mid \alpha_3\mid}^2}{\Delta}$ and
              $\Sigma=\Sigma_2+\Sigma_3$;
\end{itemize}
and
$$U=\left(\begin{array}{ccccc}
u & -\alpha_2u &  -\alpha_3u & 0 &0\\
-\overline{a}_2u & \overline{a}_2\alpha_2u &  \overline{a}_2\alpha_3u & 0 &0\\
-\overline{a}_3u & \overline{a}_3\alpha_2u &  \overline{a}_3\alpha_3u & 0 &0\\
0 & 0 & 0 & 0 & 0\\
0 & 0 & 0 & 0 & 0\\
\end{array}\right).
$$
Matrices $P$, $M$ and $Z$ have the same form of $Q$, $N$ and $U$
respectively, with $p$, $m$, $z$, $a_i$   and $b_j$ in place of
$q$, $n$, $u$, $\alpha_i$ and $\beta_j$, where $i=2,3$ and
$j=4,5$;  $A_i$, $B_i$, $C$, $D$ take the  place of $\Lambda_i$,
$\Sigma_i$, $\Gamma$, $\Delta$ and are defined in analogous
manner;  moreover, $V=\overline{U}^t$ and $W=\overline{Z}^t$. We
notice that $a_i,b_j,l_j,\alpha_i,\beta_j,\lambda_j$, with $i=2,3$
and $j=4,5$, are constant complex numbers arising from the linear
dependence among the vector-components of $\bf{x}_i$ and
$\bf{y}_j$, where $i,j=1,\ldots,5$, as we shall see in the next
section.\par In order to solve  the problem, such solution must be
idempotent. By imposing idempotence we find that
\begin{itemize}
      \item[1.]$\frac{A_2}{1+{\mid a_2 \mid}^2+{\mid a_3 \mid}^2}<p<\frac{A_2+1}{1+{\mid a_2 \mid}^2+{\mid a_3 \mid}^2}$
      \item[2.]$\frac{A_2+B_3}{1+{\mid a_2 \mid}^2+{\mid a_3 \mid}^2}<m<\frac{A_2+B_3+1}{1+{\mid a_2 \mid}^2+{\mid a_3 \mid}^2}$
      \item[3.]$u=e^{i\theta_1}\sqrt\frac{\left(p-\frac{1}{C}\right)\left(1-2A_3\right)-
               {\left(p-\frac{1}{C}\right)}^2\left(1+{\mid a_2 \mid}^2+{\mid a_3
               \mid}^2\right)+ A_3\frac{{\mid \beta_4 \mid}^2+{\mid \beta_5
               \mid}^2}{C}}{1+{\mid \alpha_2 \mid}^2+{\mid \alpha_3 \mid}^2}$
      \item[4.]$z=e^{i\theta_2}\sqrt\frac{\left(m-\frac{1}{C}\right)\left(1-2\left(A_3-B_3\right)\right)-
               {\left(m-\frac{1}{C}\right)}^2\left(1+{\mid a_2 \mid}^2+{\mid a_3
               \mid}^2\right)-\frac{(B_3-A_3)^2+ (B_3-A_3)}{1+{\mid a_2
               \mid}^2+{\mid a_3 \mid}^2}}{1+{\mid \alpha_2 \mid}^2+{\mid
               \alpha_3 \mid}^2}$
      \item[5.]$q=\frac{1+\Lambda_2+A_2-p\left(1+{\mid a_2 \mid}^2+{\mid a_3 \mid}^2\right)}{1+{\mid \alpha_2 \mid}^2+
               {\mid \alpha_3 \mid}^2}$
      \item[6.]$n=\frac{1+A_2+B_3+\Lambda_2+\Sigma_3-m\left(1+{\mid a_2 \mid}^2+
               {\mid a_3 \mid}^2\right)}{1+{\mid \alpha_2 \mid}^2+{\mid \alpha_3
               \mid}^2}$
\end{itemize}
where $\theta_1$ and $\theta_2$ are real numbers. Our family of
solutions completely solves the problem if it satisfies $[G,L]\neq
0$.\\
Therefore, for every real number $\frac{A_2}{1+{\mid a_2 \mid}^2+{\mid a_3 \mid}^2}<p<\frac{A_2+1}{1+{\mid a_2 \mid}^2+{\mid a_3 \mid}^2}$, every  real number $\frac{A_2+B_3}{1+{\mid a_2 \mid}^2+{\mid a_3 \mid}^2}<m<\frac{A_2+B_3+1}{1+{\mid a_2 \mid}^2+{\mid a_3 \mid}^2}$, every $a_2,a_3,b_4,b_5,l_5,\alpha_2,\alpha_3,\beta_4,\beta_5,\lambda_5\in\bf C$ we have a solution of \emph{(P)}. Since $G_I$ and $L_I$ are projection operators and no constraint is imposed to their traces, i.e. to $Rank(G_I)$  and $Rank(L_I)$, these parameters are not all independent.\\
For instance, the following solution of \emph{(P)}
$$
G=\left(\scriptstyle{
\begin{array}{cccccccccc}
 \frac{11}{72} & -\frac{1}{36} & -\frac{11}{72} & -\frac{1}{8} & -\frac{1}{8} & \frac{\sqrt{2}}{9} & -\frac{\sqrt{2}}{9} & -\frac{\sqrt{2}}{9} &
0 & 0 \\[10pt]
 -\frac{1}{36} & \frac{5}{18} & \frac{1}{36} & -\frac{1}{4} & -\frac{1}{4} & -\frac{\sqrt{2}}{9} & \frac{\sqrt{2}}{9} & \frac{\sqrt{2}}{9} & 0 &
0 \\[10pt]
 -\frac{11}{72} & \frac{1}{36} & \frac{11}{72} & \frac{1}{8} & \frac{1}{8} & -\frac{\sqrt{2}}{9} & \frac{\sqrt{2}}{9} & \frac{\sqrt{2}}{9} & 0 &
0 \\[10pt]
 -\frac{1}{8} & -\frac{1}{4} & \frac{1}{8} & \frac{3}{8} & \frac{3}{8} & 0 & 0 & 0 & 0 & 0 \\[10pt]
 -\frac{1}{8} & -\frac{1}{4} & \frac{1}{8} & \frac{3}{8} & \frac{3}{8} & 0 & 0 & 0 & 0 & 0 \\[10pt]
 \frac{\sqrt{2}}{9} & -\frac{\sqrt{2}}{9} & -\frac{\sqrt{2}}{9} & 0 & 0 & \frac{19}{72} & -\frac{5}{36} & -\frac{19}{72} & -\frac{1}{8} & -\frac{1}{8}
\\[10pt]
 -\frac{\sqrt{2}}{9} & \frac{\sqrt{2}}{9} & \frac{\sqrt{2}}{9} & 0 & 0 & -\frac{5}{36} & \frac{7}{18} & \frac{5}{36} & -\frac{1}{4} & -\frac{1}{4}
\\[10pt]
 -\frac{\sqrt{2}}{9} & \frac{\sqrt{2}}{9} & \frac{\sqrt{2}}{9} & 0 & 0 & -\frac{19}{72} & \frac{5}{36} & \frac{19}{72} & \frac{1}{8} & \frac{1}{8}
\\[10pt]
 0 & 0 & 0 & 0 & 0 & -\frac{1}{8} & -\frac{1}{4} & \frac{1}{8} & \frac{3}{8} & \frac{3}{8} \\[10pt]
 0 & 0 & 0 & 0 & 0 & -\frac{1}{8} & -\frac{1}{4} & \frac{1}{8} & \frac{3}{8} & \frac{3}{8}
\end{array}}
\right)$$

$$L=\left(\scriptscriptstyle{
\begin{array}{cccccccccc}
 \frac{67}{456} & -\frac{5}{228} & \frac{5}{456} & -\frac{3}{152} & -\frac{43}{152} & \frac{4}{19 \sqrt{3}} & -\frac{4}{19 \sqrt{3}} & -\frac{4}{19
\sqrt{3}} & 0 & 0 \\[10pt]
 -\frac{5}{228} & \frac{31}{114} & -\frac{31}{228} & -\frac{27}{76} & -\frac{7}{76} & -\frac{4}{19 \sqrt{3}} & \frac{4}{19 \sqrt{3}} & \frac{4}{19
\sqrt{3}} & 0 & 0 \\[10pt]
 \frac{5}{456} & -\frac{31}{228} & \frac{139}{456} & \frac{51}{152} & -\frac{29}{152} & -\frac{4}{19 \sqrt{3}} & \frac{4}{19 \sqrt{3}} & \frac{4}{19
\sqrt{3}} & 0 & 0 \\[10pt]
 -\frac{3}{152} & -\frac{27}{76} & \frac{51}{152} & \frac{89}{152} & \frac{9}{152} & 0 & 0 & 0 & 0 & 0 \\[10pt]
 -\frac{43}{152} & -\frac{7}{76} & -\frac{29}{152} & \frac{9}{152} & \frac{129}{152} & 0 & 0 & 0 & 0 & 0 \\[10pt]
 \frac{4}{19 \sqrt{3}} & -\frac{4}{19 \sqrt{3}} & -\frac{4}{19 \sqrt{3}} & 0 & 0 & \frac{3}{8} & -\frac{1}{4} & -\frac{33}{152} & -\frac{3}{152}
& -\frac{43}{152} \\[10pt]
 -\frac{4}{19 \sqrt{3}} & \frac{4}{19 \sqrt{3}} & \frac{4}{19 \sqrt{3}} & 0 & 0 & -\frac{1}{4} & \frac{1}{2} & \frac{7}{76} & -\frac{27}{76} & -\frac{7}{76}
\\[10pt]
 -\frac{4}{19 \sqrt{3}} & \frac{4}{19 \sqrt{3}} & \frac{4}{19 \sqrt{3}} & 0 & 0 & -\frac{33}{152} & \frac{7}{76} & \frac{81}{152} & \frac{51}{152}
& -\frac{29}{152} \\[10pt]
 0 & 0 & 0 & 0 & 0 & -\frac{3}{152} & -\frac{27}{76} & \frac{51}{152} & \frac{89}{152} & \frac{9}{152} \\[10pt]
 0 & 0 & 0 & 0 & 0 & -\frac{43}{152} & -\frac{7}{76} & -\frac{29}{152} & \frac{9}{152} & \frac{129}{152}
\end{array}}
\right)$$
$$\Psi={({\bf x}_1,{\bf x}_2,{\bf x}_3,{\bf x}_4,{\bf x}_5;{\bf y}_1,{\bf y}_2,{\bf y}_3,{\bf y}_4,{\bf y}_5)}^t
$$
where
\begin{equation}\begin{array}{l}\nonumber
{\bf x}_1={\left(-\frac{1}{3}{\bf a}_5,{\bf 0},-\frac{1}{3}{\bf c}_5,{\bf 0},\frac{1}{3}{\bf e}_4+2{\bf e}_5,{\bf 0},{\bf 0},{\bf 0},\right)}^t\\[10pt]
{\bf x}_2={\left(-\frac{2}{3}{\bf a}_5,{\bf 0},\frac{1}{3}{\bf c}_5,{\bf 0},{\bf e}_4+{\bf e}_5,{\bf 0},{\bf 0},{\bf 0},\right)}^t\\[10pt]
{\bf x}_3={\left(\frac{1}{3}{\bf a}_5,{\bf 0},-\frac{2}{3}{\bf c}_5,{\bf 0},-\frac{2}{3}{\bf e}_4+{\bf e}_5,{\bf 0},{\bf 0},{\bf 0},\right)}^t\\[10pt]
{\bf x}_4={\left({\bf a}_5,{\bf 0},-\frac{2}{3}{\bf c}_5,{\bf 0},{\bf e}_4,{\bf 0},{\bf 0},{\bf 0},\right)}^t\\[10pt]
{\bf x}_5={\left({\bf a}_5,{\bf 0},{\bf c}_5,{\bf 0},{\bf e}_5,{\bf 0},{\bf 0},{\bf 0},\right)}^t\\[10pt]
{\bf y}_1={\left({\bf 0},{\bf 0},{\bf 0},-\frac{1}{3}\boldsymbol{\delta}_5,{\bf 0},{\bf 0},-\frac{1}{3}\boldsymbol{\eta}_5,\frac{1}{3}\boldsymbol{\theta}_4+2\boldsymbol{\theta}_5\right)}^t\\[10pt]
{\bf y}_2={\left({\bf 0},{\bf 0},{\bf 0},-\frac{2}{3}\boldsymbol{\delta}_5,{\bf 0},{\bf 0},\frac{1}{3}\boldsymbol{\eta}_5,\boldsymbol{\theta}_4+\boldsymbol{\theta}_5\right)}^t\\[10pt]
{\bf y}_3={\left({\bf 0},{\bf 0},{\bf 0},\frac{1}{3}\boldsymbol{\delta}_5,{\bf 0},{\bf 0},-\frac{2}{3}\boldsymbol{\eta}_5,-\frac{2}{3}\boldsymbol{\theta}_4+\boldsymbol{\theta}_5\right)}^t\\[10pt]
{\bf y}_4={\left({\bf 0},{\bf 0},{\bf 0},\boldsymbol{\delta}_5,{\bf 0},{\bf 0},-\frac{2}{3}\boldsymbol{\eta}_5,\boldsymbol{\theta}_4\right)}^t\\[10pt]
{\bf y}_5={\left({\bf 0},{\bf 0},{\bf 0},\boldsymbol{\delta}_5,{\bf 0},{\bf 0},\boldsymbol{\eta}_5,\boldsymbol{\theta}_5\right)}^t\\
\end{array}\end{equation}
is obtained in corrispondence  with the particular choice $a_2=a_3=\alpha_2=\alpha_3=1$, $b_4=b_5=\beta_4=\beta_5=1$, $\lambda_5=l_5=1$ $\theta_1=\theta_2=0$, $p=\frac{11}{72}$ and $m=\frac{67}{456}$.\\
\noindent We stress that our family of  solutions is a particular  one, attained by making particular assumptions and corresponding to correlated detections. The question whether non-correlated solutions do exist or not, if $dim({\cal H}_1)=10$, remains open.

\section{Derivation of a family of solutions}\protect\label{sec:derivation of four}
In this section we carry out the  detailed derivation of the
family of solutions of problem (${\cal P'}$) presented in previous
section. Our treatment is not at all exhaustive. Indeed, we shall
consider solutions characterized by particular conditions of
linear independence between some of their components. In our
derivation we analyze the equations involving vectors $y_i$, i.e.
(ii-B), (iv-D) in (\protect\ref{eq:sistema I}) and (ii-B'),
(iv-D') in (\protect\ref{eq:sistema II}); in order to solve them,
we make some assumptions (e.g. vectors $\left(
\begin{array}{c}
\boldsymbol{\eta}_{3}\\
\boldsymbol{\theta}_{3}\\
\end{array} \right)$, $\ldots$, $\left( \begin{array}{c}
\boldsymbol{\eta}_{n}\\
\boldsymbol{\theta}_{n}\\
\end{array} \right)$ are supposed linearly independent, as well as $\boldsymbol{\theta}_{4},\ldots,\boldsymbol{\theta}_{n}$;
furthermore, $\boldsymbol{\zeta}_j={\bf 0}$ for all
$j=1,\ldots,n$,) which lead to particular forms for vectors $y_i$
and for matrices $Q$, $U$, $N$, $Z$. Analogous results are taken
also for (i-A), (iii-C) in (\protect\ref{eq:sistema I}) and
(i-A'), (iii-C') in (\protect\ref{eq:sistema II}), which have the
same form of the previous ones; so we obtain a particular form for
vectors  $x_i$ and for matrices $P$, $V$, $M$, $W$. Then we fix
the dimension of ${\cal H}_I$, $dim({\cal H}_I)=10$; by fixing the
value of parameters (see (\protect\ref{eq:choice})) we obtain the
particular family of solutions of problem $({\cal P'})$, singled
out in previous section.
%, which show a correlation between $L$ and $G$.
\subsection{General constraints for $\Psi$ and $G$}
Equations in (ii-B) imply $u_{j1}{\bf y}_1+\cdots+u_{jn}{\bf
y}_n=0$. Therefore, since  $U\neq {\bf 0}$, vectors ${\bf
y}_1,\ldots, {\bf y}_n$ must be linearly dependent. Let us suppose
${\bf y}_1=\alpha_{2}{\bf y}_{2}+\cdots+\alpha_{n}{\bf y}_{n}$. By
using this relation in  (iv-D) we get
\begin{equation}\protect\label{eq:(iv-D)1}
\left\{\begin{array}{lll}
Q_{j2}\boldsymbol{\delta}_{2}+\cdots+Q_{jn}\boldsymbol{\delta}_{n}=\boldsymbol{\delta}_j\\[10pt]
Q_{j2}\boldsymbol{\zeta}_{2}+\cdots+Q_{jn}\boldsymbol{\zeta}_{n}=\boldsymbol{\zeta}_j\\[10pt]
{Q_{j2}} \left( \begin{array}{c}
\boldsymbol{\eta}_{2}\\
\boldsymbol{\theta}_{2}\\
\end{array} \right)
+\cdots+{Q_{jn}} \left( \begin{array}{ll}
\boldsymbol{\eta}_{n}\\
\boldsymbol{\theta}_{n}\\
\end{array} \right)=
\left( \begin{array}{ll}
{\bf 0} \\
{\bf 0}
\end{array} \right).
\end{array}
\right.
\end{equation}
where $Q_{jk}=q_{jk}+\alpha_{k}q_{j1}$, for all $k=2,\ldots, n$.
\par In the third equation of (\protect\ref{eq:(iv-D)1}), if we
suppose $\left( \begin{array}{c}
\boldsymbol{\eta}_{2}\\
\boldsymbol{\theta}_{2}\\
\end{array} \right),\ldots,\left( \begin{array}{c}
\boldsymbol{\eta}_{n}\\
\boldsymbol{\theta}_{n}\\
\end{array} \right)$ linearly independent, then $Q_{j2}=\ldots=Q_{jn}=0$, which implies $\boldsymbol{\delta}_j=\boldsymbol{\zeta}_j={\bf 0}$, for all
$j=1,\ldots,n$, in the first and second equation of
(\protect\ref{eq:(iv-D)1}). A similar reasoning for  (i-A) leads
to ${\bf a}_j={\bf b}_j={\bf 0}$, for all $j=1,\ldots,n$; as a
consequence ${\bf x}_j={({\bf 0},{\bf 0},{\bf c}_j,{\bf 0},{\bf
e}_j,{\bf 0},{\bf 0},{\bf 0})}^t$ and ${\bf y}_j={({\bf 0},{\bf
0},{\bf 0},{\bf 0},{\bf 0},{\bf
0},\boldsymbol{\eta}_j,\boldsymbol{\theta}_j)}^t$ so that, if a
solution exists, then $Y\Psi=0$, i.e. it would be correspond to
the uninteresting case excluded by (C.10). If we consider all
cases,  we obtain the following implications:
\begin{itemize}
      \item[a.)] $\left( \begin{array}{c}
                 {\bf c}_{2}\\
                 {\bf e}_{2}\\
                 \end{array} \right),\ldots,\left( \begin{array}{c}
                 {\bf c}_{n}\\
                 {\bf e}_{n}\\
                 \end{array} \right)$  linearly independent and $\left( \begin{array}{c}
                 \boldsymbol{\eta}_{2}\\
                 \boldsymbol{\theta}_{2}\\
                 \end{array} \right),\ldots,\left( \begin{array}{c}
                 \boldsymbol{\eta}_{n}\\
                 \boldsymbol{\theta}_{n}\\
                 \end{array} \right)$ linearly independent
                imply  ${\bf x}_j={({\bf 0},{\bf
                0},{\bf c}_j,{\bf 0},{\bf e}_j,{\bf 0},{\bf 0},{\bf 0})}^t$ and
                ${\bf y}_j={({\bf 0},{\bf 0},{\bf 0},{\bf 0},{\bf 0},{\bf
                0},\boldsymbol{\eta}_j,\boldsymbol{\theta}_j)}^t$.
      \item[b.)]$\left( \begin{array}{c}
                 {\bf c}_{2}\\
                 {\bf e}_{2}\\
                 \end{array} \right),\ldots,\left( \begin{array}{c}
                 {\bf c}_{n}\\
                 {\bf e}_{n}\\
                 \end{array} \right)$  linearly independent and $\left( \begin{array}{c}
                 \boldsymbol{\eta}_{2}\\
                 \boldsymbol{\theta}_{2}\\
                 \end{array} \right),\ldots,\left( \begin{array}{c}
                 \boldsymbol{\eta}_{n}\\
                 \boldsymbol{\theta}_{n}\\
                 \end{array} \right)$ linearly dependent
                imply ${\bf x}_j={({\bf 0},{\bf 0},{\bf c}_j,{\bf 0},{\bf e}_j,{\bf
                0},{\bf 0},{\bf 0})}^t$ and ${\bf y}_j={({\bf 0},{\bf 0},{\bf
                0},\boldsymbol{\delta}_j,{\bf
                0},\boldsymbol{\zeta}_j,\boldsymbol{\eta}_j,\boldsymbol{\theta}_j)}^t$;
      \item[c.)] $\left( \begin{array}{c}
                 {\bf c}_{2}\\
                 {\bf e}_{2}\\
                 \end{array} \right),\ldots,\left( \begin{array}{c}
                 {\bf c}_{n}\\
                 {\bf e}_{n}\\
                 \end{array} \right)$  linearly dependent and $\left( \begin{array}{c}
                 \boldsymbol{\eta}_{2}\\
                 \boldsymbol{\theta}_{2}\\
                 \end{array} \right),\ldots,\left( \begin{array}{c}
                 \boldsymbol{\eta}_{n}\\
                 \boldsymbol{\theta}_{n}\\
                 \end{array} \right)$ linearly independent
                imply ${\bf x}_j={({\bf a}_j,{\bf
                b}_j,{\bf c}_j,{\bf 0},{\bf e}_j,{\bf 0},{\bf 0},{\bf 0})}^t$ and
                ${\bf y}_j={({\bf 0},{\bf 0},{\bf 0},{\bf 0},{\bf 0},{\bf
                0},\boldsymbol{\eta}_j,\boldsymbol{\theta}_j)}^t$;
      \item[d.)]$\left( \begin{array}{c}
                 {\bf c}_{2}\\
                 {\bf e}_{2}\\
                 \end{array} \right),\ldots,\left( \begin{array}{c}
                 {\bf c}_{n}\\
                 {\bf e}_{n}\\
                 \end{array} \right)$  linearly dependent and $\left( \begin{array}{c}
                 \boldsymbol{\eta}_{2}\\
                 \boldsymbol{\theta}_{2}\\
                 \end{array} \right),\ldots,\left( \begin{array}{c}
                 \boldsymbol{\eta}_{n}\\
                 \boldsymbol{\theta}_{n}\\
                 \end{array} \right)$ linearly dependent
                imply ${\bf x}_j={({\bf a}_j,{\bf b}_j,{\bf c}_j,{\bf 0},{\bf e}_j,{\bf
                0},{\bf 0},{\bf 0})}^t$  and ${\bf y}_j={({\bf 0},{\bf 0},{\bf
                0},\boldsymbol{\delta}_j,{\bf
                0},\boldsymbol{\zeta}_j,\boldsymbol{\eta}_j,\boldsymbol{\theta}_j)}^t$.
\end{itemize}
We shall search solutions for cases (b), (c) and (d), since in
case (a) meaningful solutions cannot exist.\par

\vspace{10pt} \centerline{\bf Cases (b) and (c)}\vspace{10pt}
According to (b), conditions (\protect\ref{eq:GPsi}) and
(\protect\ref{eq:LPsi}) imply equation $TW\Psi={\bf 0}$ holds,
which is equivalent to say that each time a particle is sorted by
$W$ than it is certainly not sorted by $T$; therefore, for all
eventual solutions corresponding to this case, property $L$ must
be correlated with WS property $E$.
\par In case (c) $T'W'Y\Psi=0$ holds; this equation expresses the impossibility that
two probabilities are zero and the remaining is 1, for the
occurrence of $T'$, $W'$ and $Y$; in any case, properties $L$ and
$G$ are correlated with WS property $E$.\par

%so that no correlated or meaningless solution immediately follows.
\vspace{10pt}\centerline{\bf Case (d)} \vspace{10pt}
%Now we investigate case (d.).
No correlated or meaningless solution immediately follows from
case (d). We restrict our search by working with the equations in
(ii-B), (iv-D) of (\protect\ref{eq:sistema I}), rather than with
all of them. Since (i-A), (iii-C) in (\protect\ref{eq:sistema I})
are formally identical to (ii-B), (iv-D) of
(\protect\ref{eq:sistema I}), we can extend to them the results
found for (ii-B), (iv-D). \par Our task is more simple if we
search solutions corresponding to particular state vectors $\Psi$
satisfying (d); for this reason, among vectors $\left(
\begin{array}{c}
                 \boldsymbol{\eta}_{2}\\
                 \boldsymbol{\theta}_{2}\\
                 \end{array} \right),\ldots,\left( \begin{array}{c}
                 \boldsymbol{\eta}_{n}\\
                 \boldsymbol{\theta}_{n}\\
                 \end{array} \right)$,
we suppose that only $\left( \begin{array}{c}
                 \boldsymbol{\eta}_{2}\\
                 \boldsymbol{\theta}_{2}\\
                 \end{array} \right)$
is a linear combination of the remaining ones; let
$\beta_3,\ldots,\beta_n$ be complex numbers such that
\begin{equation}\protect\label{eq:eta theta}\left( \begin{array}{c}
\boldsymbol{\eta}_{2}\\
\boldsymbol{\theta}_{2}\\
\end{array} \right)=\beta_{3}\left( \begin{array}{c}
\boldsymbol{\eta}_{3}\\
\boldsymbol{\theta}_{3}\\
\end{array} \right)+\ldots+\beta_{n}\left( \begin{array}{c}
\boldsymbol{\eta}_{n}\\
\boldsymbol{\theta}_{n}\\
\end{array} \right),\end{equation}
where $\left( \begin{array}{c}
\boldsymbol{\eta}_{3}\\
\boldsymbol{\theta}_{3}\\
\end{array} \right),\ldots,\left( \begin{array}{c}
\boldsymbol{\eta}_{n}\\
\boldsymbol{\theta}_{n}\\
\end{array} \right)$
are supposed linearly independent. Hence,  (ii-B) and (iv-D) yield
\begin{equation}\protect\label{eq:(ii-B)1}
\left\{\begin{array}{ll} U_{j2}\left( \begin{array}{c}
\boldsymbol{\delta}_{2}\\
\boldsymbol{\zeta}_{2}\\
\end{array} \right)+\ldots+U_{jn}\left( \begin{array}{c}
\boldsymbol{\delta}_{n}\\
\boldsymbol{\zeta}_{n}\\
\end{array}\right)=\left( \begin{array}{c}
{\bf 0}\\
{\bf 0}\\
\end{array} \right)\\[10pt]
\left(U_{j2}\beta_{3}+U_{j3}\right)\left( \begin{array}{c}
\boldsymbol{\eta}_{3}\\
\boldsymbol{\theta}_{3}\\
\end{array} \right)+\ldots+\left(U_{j2}\beta_{n}+U_{jn}\right)\left( \begin{array}{c}
\boldsymbol{\eta}_{n}\\
\boldsymbol{\theta}_{n}\\
\end{array} \right)=\left( \begin{array}{c}
{\bf 0}\\
{\bf 0}\\
\end{array} \right)
\end{array} \right.
\end{equation}
where $U_{jk}=u_{jk}+\alpha_k u_{j1}$, for all $k=2,\ldots,n$, and
\begin{equation}\protect\label{eq:(IV-D)2}
\left\{\begin{array}{ll} Q_{j2}\left( \begin{array}{c}
\boldsymbol{\delta}_{2}\\
\boldsymbol{\zeta}_{2}\\
\end{array} \right)+
\ldots+Q_{jn}\left( \begin{array}{c}
\boldsymbol{\delta}_{n}\\
\boldsymbol{\zeta}_{n}\\
\end{array} \right)=\left( \begin{array}{c}
\boldsymbol{\delta}_j\\
\boldsymbol{\zeta}_j\\
\end{array} \right)\\[10pt]
\left(Q_{j2}\beta_{3}+Q_{j3}\right)\left( \begin{array}{c}
\boldsymbol{\eta}_{3}\\
\boldsymbol{\theta}_{3}\\
\end{array} \right)+\ldots+\left(Q_{j2}\beta_{n}+Q_{jn}\right)\left( \begin{array}{c}
\boldsymbol{\eta}_{n}\\
\boldsymbol{\theta}_{n}\\
\end{array} \right)=\left( \begin{array}{c}
{\bf 0}\\
{\bf 0}
\end{array} \right)
\end{array} \right.\end{equation}
where $Q_{jk}=q_{jk}+\alpha_k q_{j1}$, for all $k=2,\ldots,n$. The
linear independence among vectors in the second equation
 of (\protect\ref{eq:(ii-B)1}) and
(\protect\ref{eq:(IV-D)2}) implies that
\begin{equation}\protect\label{eq:U}
U_{j2}\beta_{3}+U_{j3}=\ldots=U_{j2}\beta_{n}+U_{jn}=0,
\end{equation}
\begin{equation}\protect\label{eq:Q}
Q_{j2}\beta_{3}+Q_{j3}=\ldots=Q_{j2}\beta_{n}+Q_{jn}=0
\end{equation}
respectively. Then (\protect\ref{eq:(ii-B)1}) and
(\protect\ref{eq:(IV-D)2}) become
\begin{equation}\protect\label{eq:(ii-B)2}\left\{\begin{array}{ll}
U_{j2}\left[\left( \begin{array}{c}
\boldsymbol{\delta}_{2}\\
\boldsymbol{\zeta}_{2}\\
\end{array} \right)-\beta_{3}\left( \begin{array}{c}
\boldsymbol{\delta}_{3}\\
\boldsymbol{\zeta}_{3}\\
\end{array} \right)-
\ldots-\beta_{n}\left( \begin{array}{c}
\boldsymbol{\delta}_{n}\\
\boldsymbol{\zeta}_{n}\\
\end{array} \right)\right]=\left( \begin{array}{c}
{\bf 0}\\
{\bf 0}\\
\end{array} \right) \\[10pt]
U_{j2}\beta_{3}+U_{j3}=\ldots=U_{j2}\beta_{n}+U_{jn}=0;
\end{array} \right.\end{equation}
\begin{equation}\protect\label{eq:(IV-D)3}
\left\{\begin{array}{ll} Q_{j2}\left[\left( \begin{array}{c}
\boldsymbol{\delta}_{2}\\
\boldsymbol{\zeta}_{2}\\
\end{array} \right)-\beta_{3}\left( \begin{array}{c}
\boldsymbol{\delta}_{3}\\
\boldsymbol{\zeta}_{3}\\
\end{array} \right)-
\ldots-\beta_{n}\left( \begin{array}{c}
\boldsymbol{\delta}_{n}\\
\boldsymbol{\zeta}_{n}\\
\end{array} \right)\right]=\left( \begin{array}{c}
\boldsymbol{\delta}_j\\
\boldsymbol{\zeta}_j\\
\end{array} \right)\\[10pt]
Q_{j2}\beta_{3}+Q_{j3}=\ldots=Q_{j2}\beta_{n}+Q_{jn}=0.
\end{array} \right.\end{equation}
First equation in (\protect\ref{eq:(ii-B)2}) is satisfied if one
of the two factor is zero; however, if we suppose
\begin{equation*}\left( \begin{array}{c}
\boldsymbol{\delta}_{2}\\
\boldsymbol{\zeta}_{2}\\
\end{array} \right)=\beta_{3}\left( \begin{array}{c}
\boldsymbol{\delta}_{3}\\
\boldsymbol{\zeta}_{3}\\
\end{array} \right)+
\ldots+\beta_{n}\left( \begin{array}{c}
\boldsymbol{\delta}_{n}\\
\boldsymbol{\zeta}_{n}\\
\end{array} \right).\end{equation*}
(\protect\ref{eq:(IV-D)3})  implies
$\boldsymbol{\delta}_j=\boldsymbol{\zeta}_j={\bf 0}$, for all
$j=1,\ldots,n$; since we are taking analogous results for (i-A)
and (iii-C), then ${\bf a}_j={\bf b}_j={\bf 0}$ for all
$j=1,\ldots,n$. As a consequence, ${\bf x}_j={({\bf 0},{\bf
                0},{\bf c}_j,{\bf 0},{\bf e}_j,{\bf 0},{\bf 0},{\bf 0})}^t$ and
                ${\bf y}_j={({\bf 0},{\bf 0},{\bf 0},{\bf 0},{\bf 0},{\bf
                0},\boldsymbol{\eta}_j,\boldsymbol{\theta}_j)}^t$,
which lead to meaningless solutions (case (a)). Hence, $U_{j2}=0$
follows from (\protect\ref{eq:(ii-B)2}).\par In
(\protect\ref{eq:(IV-D)3}) we can suppose that vectors $\left(
\begin{array}{c}
\boldsymbol{\delta}_{2}\\
\boldsymbol{\zeta}_{2}\\
\end{array} \right),\ldots, \left( \begin{array}{c}
\boldsymbol{\delta}_{n}\\
\boldsymbol{\zeta}_{n}\\
\end{array} \right)$
are either linearly independent or linearly dependent (with
coefficients different from ${ b}_3,\ldots,{ b}_n$). In case they
are independent, first equation in (\protect\ref{eq:(IV-D)3})
yields
\begin{equation}\protect\label{eq:(IV-D)4}
\left\{\begin{array}{lllll}
Q_{n 2}=0\\
{ \beta}_{3}Q_{n 2}=0\\
\vdots\\
{ \beta}_{n-1}Q_{n 2}=0\\
{ \beta}_{n}Q_{n 2}=1
\end{array} \right.\end{equation}
in correspondence with $j=n$, which has no solution. So we can
suppose the existence of complex numbers, $c_3,\ldots,c_n$ such
that
$$\left( \begin{array}{c}
\boldsymbol{\delta}_{2}\\
\boldsymbol{\zeta}_{2}\\
\end{array} \right)=c_{3}\left( \begin{array}{c}
\boldsymbol{\delta}_{3}\\
\boldsymbol{\zeta}_{3}\\
\end{array} \right)+
\ldots+c_{n}\left( \begin{array}{c}
\boldsymbol{\delta}_{n}\\
\boldsymbol{\zeta}_{n}\\
\end{array} \right).$$
Again, with a reasoning similar to the previous one, carried out
for $j=n$ and $j=n-1$, no solution is found for
(\protect\ref{eq:(IV-D)3}). The same conclusion can be drawn if
just two vectors among $\left( \begin{array}{c}
\boldsymbol{\delta}_{2}\\
\boldsymbol{\zeta}_{2}\\
\end{array} \right),\ldots,\left( \begin{array}{c}
\boldsymbol{\delta}_{n}\\
\boldsymbol{\zeta}_{n}\\
\end{array} \right)$
are linearly independent, so  we may conclude that only one is
independent, say the last. Hence complex numbers
${\gamma}_{2},\ldots,\gamma_{n-1}$ must exist such that
\begin{equation}\protect\label{eq:delta zeta}
\left( \begin{array}{c}
\boldsymbol{\delta}_{2}\\
\boldsymbol{\zeta}_{2}\\
\end{array} \right)={\gamma}_{2}\left( \begin{array}{c}
\boldsymbol{\delta}_{n}\\
\boldsymbol{\zeta}_{n}\\
\end{array} \right),\ldots,\left( \begin{array}{c}
\boldsymbol{\delta}_{n-1}\\
\boldsymbol{\zeta}_{n-1}\\
\end{array} \right)={\gamma}_{n-1}\left( \begin{array}{c}
\boldsymbol{\delta}_{n}\\
\boldsymbol{\zeta}_{n}\\
\end{array} \right).\end{equation}
As a consequence (\protect\ref{eq:(IV-D)3}) can be written as
\begin{equation}\protect\label{eq:systems}\left\{\begin{array}{lll}
{\cal
Q}_{j2}\left[\gamma_{2}-\beta_{3}\gamma_{3}-\ldots-\beta_{n-1}\gamma_{n-1}-\beta_{n}\right]\left(
\begin{array}{c}
\boldsymbol{\delta}_{n}\\
\boldsymbol{\zeta}_{n}\\
\end{array} \right)=\left( \begin{array}{c}
\boldsymbol{\delta}_{j}\\
\boldsymbol{\zeta}_{j}\\
\end{array} \right)\\[10pt]
Q_{j2}\beta_{3}+Q_{j3}=\ldots=Q_{j2}\beta_{n}+Q_{jn}=0.
\end{array} \right.
\end{equation}

\subsection{General constraints for $\Psi$ and $L$}
Equations in (ii-B') have the same form of those in (ii-B); hence,
following  (\protect\ref{eq:eta
theta})-(\protect\ref{eq:(ii-B)2}), conclusions  drawn for (ii-B)
continue to hold for (ii-B').% by means of the substitution
%$z_{ik}=u_{ik}$.
\par Again, in order to make easier our task, we restrict the
search  by working with the equations in (iv-D'); since (i-A') is
formally identical, we can extend to it analogous results.
\par In (iv-D'), as a consequence of (\protect\ref{eq:eta theta})
and (\protect\ref{eq:delta zeta}), we get
\begin{equation}\protect\label{eq:(iv-D')1}\left\{\begin{array}{ll}
\left(N_{j2}\gamma_{2}+\ldots+N_{j
n-1}\gamma_{n-1}+N_{jn}\right)\left( \begin{array}{c}
\boldsymbol{\delta}_{n}\\
\boldsymbol{\zeta}_{n}\\
\end{array} \right)=\left( \begin{array}{c}
\boldsymbol{\delta}_j\\
{\bf 0}\\
\end{array} \right)\\[10pt]
\left(N_{j2}\beta_{3}+N_{j3}\right)\left( \begin{array}{c}
\boldsymbol{\eta}_{3}\\
\boldsymbol{\theta}_{3}\\
\end{array} \right)+\ldots+\left(N_{j2}\beta_{n}+N_{jn}\right)\left( \begin{array}{c}
\boldsymbol{\eta}_{n}\\
\boldsymbol{\theta}_{n}\\
\end{array} \right)=\left( \begin{array}{c}
\boldsymbol{\eta}_j\\
{\bf 0}\\
\end{array} \right)
\end{array} \right.
\end{equation}
where $N_{jk}=n_{jk}+\alpha_k n_{j1}$. In order to solve first
equation, either $\boldsymbol{\zeta}_j={\bf 0}$, for all $j$, or
$N_{ji+1}\gamma_{ji+1}+\ldots+N_{j n-1}\gamma_{n-1}+N_{jn}=0$,
which  imply $\boldsymbol{\delta}_j={\bf 0}$, for all $j$; in any
case, solutions, if they exist, are correlated. Indeed, since we
are taking symmetrical results for (i-A'), we obtain the following
implications:
\begin{itemize}
\item[e.)] if $\boldsymbol{\delta}_j={\bf a}_j={\bf 0}$ for all $j=1,\ldots,n$ then $WY\Psi=0$,  i.e. each time
a particle is sorted by $Y$ than it is certainly not sorted by
$W$; therefore, for all eventual solutions corresponding to this
case, property $L$ must be correlated with the incompatible
property $G$;
\item[f.)] if $\boldsymbol{\zeta}_j={\bf b}_j={\bf 0}$ for all $j=1,\ldots,n$ then $W'Y\Psi=0$, i.e. each time
a particle is sorted by $Y$ than it is certainly not sorted by
$W'$; therefore, for all eventual solutions corresponding to this
case, property $L$ must be correlated with the incompatible
property $G$.
\end{itemize}
Let us suppose $\boldsymbol{\zeta_j}={\bf 0}$, for all $j$.\par
Equations in  (\protect\ref{eq:(iv-D')1}) can also be written as
\begin{equation}\protect\label{eq:(iv-D')2}
\left\{\begin{array}{lll}
\left(N_{j2}\gamma_{j2}+\ldots+N_{j n-1}\gamma_{n-1}+N_{jn}\right)\boldsymbol{\delta}_{n}=\boldsymbol{\delta}_j\\[10pt]
{\cal N}_{j 3}\boldsymbol{\eta}_{3}+\ldots+{\cal N}_{j n}\boldsymbol{\eta}_{n}=\boldsymbol{\eta}_j\\[10pt]
{\cal N}_{j 3}\boldsymbol{\theta}_{3}+\ldots+{\cal N}_{j n}\boldsymbol{\theta}_{n}={\bf 0}\\
\end{array} \right.\end{equation}
where ${\cal N}_{j k}=N_{j2}\beta_{k}+N_{jk}$, for all
$k=3,\ldots,n$. In the last two equations of
(\protect\ref{eq:(iv-D')2}), a reasoning similar to that carried
out for (iv-D) (see equations
(\protect\ref{eq:(iv-D)1})-(\protect\ref{eq:delta zeta})) implies
the existence of coefficients $\lambda_4,\ldots,\lambda_n$ and
$\mu_{3},\ldots, \mu_{n-1}$ such that
\begin{equation}\protect\label{eq:theta eta}
\begin{array}{ll}
\boldsymbol{\theta}_{3}=\lambda_{4}\boldsymbol{\theta}_{4}+\ldots+\lambda_{n}\boldsymbol{\theta}_{n}\\[10pt]
\boldsymbol{\eta}_{k}=\mu_{k}\boldsymbol{\eta}_{n} \quad\forall
k=3,\ldots, n-1
\end{array}\end{equation}
where we have supposed
$\boldsymbol{\theta}_{4},\ldots,\boldsymbol{\theta}_{n}$ linearly
independent. Such an independence implies
\begin{equation}\protect\label{eq:N}
{\cal N}_{j 3}\lambda_4+{\cal N}_{j4}=\ldots={\cal N}_{j
3}\lambda_n+{\cal N}_{j n}=0,
\end{equation}
in the last equation of (\protect\ref{eq:(iv-D')2}). Hence,
(iv-D') can be written as
\begin{equation}\protect\label{eq:systems}\left\{\begin{array}{lll}
\left(N_{j2}\gamma_{2}+\ldots+N_{j n-1}\gamma_{n-1}+N_{jn}\right)\boldsymbol{\delta}_{n}=\boldsymbol{\delta}_j\\[10pt]
{\cal N}_{j
3}\left[\mu_{3}-\lambda_{4}\mu_{4}-\ldots-\lambda_{n-1}\mu_{n-1}-\lambda_{n}\right]\boldsymbol{\eta}_{n}=\boldsymbol{\eta}_j\\[10pt]
{\cal N}_{j 3}\lambda_4+{\cal N}_{j4}=\ldots={\cal N}_{j
3}\lambda_n+{\cal N}_{j n}=0.
\end{array} \right.
\end{equation}

\subsection{Concrete solutions}
So far we have established some constraints in the hypothesis that
vectors $\left( \begin{array}{c}
\boldsymbol{\eta}_{3}\\
\boldsymbol{\theta}_{3}\\
\end{array} \right)$, $\ldots$, $\left( \begin{array}{c}
\boldsymbol{\eta}_{n}\\
\boldsymbol{\theta}_{n}\\
\end{array} \right)$   are linearly independent, as well as $\boldsymbol{\theta}_{4},\ldots,\boldsymbol{\theta}_{n}$;
furthermore, $\boldsymbol{\zeta}_j={\bf 0}$ for all
$j=1,\ldots,n$, independently of the ranks of matrices  $U$, $Q$,
$Z$, $N$ and
$A_i$, with $i=1\ldots,8$, and therefore of the dimensions of spaces ${\cal H}_I$ and ${\cal H}_{II}$.\\
Now we fix $dim({\cal H}_I)=10$, hence $n,j, k\in\lbrace
1,\ldots,5\rbrace$, and we shall see that concrete solutions
exist. Our task is made easier if we search for solutions
corresponding to a particular state vector $\Psi$ such that
\begin{equation}
\begin{array}{ll}\protect\label{eq:choice}
\gamma_3=\lambda_4{\gamma}_4+\lambda_5,\quad &\alpha_4=\alpha_5=\beta_3=0\\[10pt]
c_3=l_4c_4+l_5,\quad &a_4=a_5=b_3=0.
\end{array}
\end{equation}
where the coefficients that appear in the second line are those
corresponding to the vectors ${\bf x}_i$ ($i=1,\ldots,10$).
Conditions (\protect\ref{eq:choice}) and (\protect\ref{eq:U}),
together with $U_{j2}=0$ (arising from
(\protect\ref{eq:(ii-B)2})), imply that $U$ has the following form
\begin{equation}
U=\left(\begin{array}{ccccc}
u_{11} &  -\alpha_2u_{11}& -\alpha_3u_{11} &  0& 0\\
. & . & . & . & .\\
u_{j1} &  -\alpha_2u_{j1}& -\alpha_3u_{j1} &  0& 0\\
. & . & . & . & .\\
\end{array}\right);
\end{equation}
the first equation in (\protect\ref{eq:systems}), together with
(\protect\ref{eq:U}) and the choice (\protect\ref{eq:choice})
imply that $Q=(q_{ij})_{5\times 5}$ where
$q_{j3}=-\alpha_3q_{j1}$, $q_{j4}=-\beta_4(q_{j1}\alpha_2+q_{j2})$
and $q_{j5}=-\beta_5(q_{j1}\alpha_2+q_{j2})$. Similarly,
independence of $\theta_4,\ldots,\theta_n$ in
(\protect\ref{eq:(iv-D')2}) and the second equation of
(\protect\ref{eq:systems}), imply that $N=(n_{ij})_{5\times 5}$
where $n_{j2}=q_{j2}+\alpha_2(q_{j1}-n_{j1})$,
$n_{j4}=-\lambda_4\alpha_3n_{j1}-\beta_4(q_{j1}\alpha_2+q_{j2})-\lambda_4n_{j3}$
and
$n_{j5}=-\lambda_5\alpha_3n_{j1}-\beta_5(q_{j1}\alpha_2+q_{j2})-\lambda_5n_{j3}$.\\
Since (ii-B') has the same form of (ii-B),   matrix $Z$ can be
obtained from $U$ by means of the substitution $u_{ik}=z_{jk}$.
Matrices $P$, $V$, $N$, $W$ have similar forms. By imposing that
$G_I$ and $L_I$ are self-adjoint matrices we find exactly matrices
$U$, $Q$ and $N$ in section \protect\ref{sec:A family of
solutions}, and moreover $V=\overline U^t$.

\section*{Acknowledgments}
The author is greateful to G. Nistic\`o for suggesting the problem and for his generous help during the preparation of this work.

% Set the ending of a LaTeX document
\end{document}